\RequirePackage{etoolbox}
\newtoggle{quantum} \toggletrue{quantum}
\iftoggle{quantum}{
    \documentclass[onecolumn,a4paper,accepted=2026-05-26]{quantumarticle}
    \pdfoutput=1
}{
    \documentclass[letterpaper,11pt]{article}
}

\iftoggle{quantum}{}{
    \usepackage{palatino}
\usepackage[paperwidth=8.5in,paperheight=11.0in,
  left=1.0in,right=1.0in,top=1.0in,bottom=1.0in,
  includefoot,heightrounded]{geometry}
  \usepackage{amsmath,amssymb,amsthm}
\setlength{\parindent}{1em}
    \setlength{\parskip}{0.5em}
}

\usepackage[dvipsnames]{xcolor}
    \usepackage{amsthm}
\usepackage{mathtools}
\usepackage{graphicx}
    
    \usepackage{todonotes}

    \usepackage{todonotes}
    \usepackage{graphicx}
    \usepackage{amsfonts}
    \usepackage{framed}
    \usepackage{amsmath}
    \usepackage{amssymb}
    \usepackage{hyperref}
    \hypersetup{
        colorlinks=true,
        citecolor=Rhodamine,
        linkcolor=Rhodamine,
        filecolor=Rhodamine,      
        urlcolor=blue,
    }

    \newcommand{\negpar}[1][-1em]{\ifvmode\else\par\fi
      {\parindent=#1\leavevmode}\ignorespaces
    }
    
    \usepackage{tabularx}
    \usepackage{caption}
    \usepackage{subcaption}

    \usepackage[]{algorithm2e}
    \usepackage{mathtools}
    \usepackage[
      backend=bibtex,
      url=false,
      eprint=false,
      sorting=nyt
    ]{biblatex}
    \usepackage[capitalize]{cleveref}
    
    \newtheorem{theorem}{Theorem}[section]

\newtheorem{definition}[theorem]{Definition}
    
    \theoremstyle{plain}

    \theoremstyle{plain}

    \theoremstyle{plain}
    \newtheorem{lemma}[theorem]{Lemma}
    
    \theoremstyle{plain}
    \newtheorem{claim}[theorem]{Claim}

    \newtheoremstyle{myremark}     {10pt}{10pt}{}{}{\scshape}{.}{.5em}{}
    \theoremstyle{myremark}
    \newtheorem{remark}{Remark}[section]
    
    \usepackage{comment}
    \usepackage[most]{tcolorbox}
    
    \newtcolorbox{red}{
    notitle,
    colback=red!30
    }
    
    \newtcolorbox{blue}{
    notitle,
    colback=blue!30
    }
    
    \newtcolorbox{green}{
    notitle,
    colback=green!30
    }

\newcommand{\poly}{\ensuremath{\mathsf{poly}}\xspace}

\newcommand{\negl}{\ensuremath{\mathsf{negl}}\xspace}

\newcommand{\crs}{\ensuremath{\mathsf{CRS}}}

\newcommand{\Gen}{\ensuremath{\mathsf{Gen}}\xspace}

\newcommand{\xor}{\oplus}
\newcommand{\tensor}{\otimes}
\newcommand{\identity}{\ensuremath{\mathbb{I}}}

\newcommand{\bits}{\{0,1\}}

\iftoggle{quantum}{}{

}

\iftoggle{quantum}{}{
    \allowdisplaybreaks[1]
}

\newcommand{\NP}{\mathsf{NP}}
\newcommand{\QMA}{\mathsf{QMA}}

  \newcommand{\cA}{\ensuremath{{\mathcal A}}\xspace}
  \newcommand{\cB}{\ensuremath{{\mathcal B}}\xspace}
  \newcommand{\cC}{\ensuremath{{\mathcal C}}\xspace}
  \newcommand{\cD}{\ensuremath{{\mathcal D}}\xspace}

  \newcommand{\cH}{\ensuremath{{\mathcal H}}\xspace}
  \newcommand{\cI}{\ensuremath{{\mathcal I}}\xspace}
  \newcommand{\cM}{\ensuremath{{\mathcal M}}\xspace}
  
  \newcommand{\cO}{\ensuremath{{\mathcal O}}\xspace}

  \newcommand{\cR}{\ensuremath{{\mathcal R}}\xspace}

  \newcommand{\cX}{\ensuremath{{\mathcal X}}\xspace}

\newcommand{\zo}{\ensuremath{{\{0,1\}}}\xspace}

\newcommand{\secpar}{\kappa}

\newcommand{\sthat}{\: : \:}
\newcommand{\cL}{\ensuremath{{\mathcal L}}\xspace}

\theoremstyle{property}
    \newtheorem{property}{Property}[section]
    \usepackage{setspace}

\newcommand{\sP}{\mathsf{P}}
\newcommand{\sV}{\mathsf{V}}

\newcommand{\Sim}{\mathsf{Sim}}

\newcommand{\accept}{\mathsf{accept}}

\newcommand{\Com}{\mathsf{Com}}
\newcommand{\DualCom}{\mathsf{DualCom}}

\newcommand{\td}{\mathsf{td}}

\newcommand{\bbN}{\mathbb{N}}
\newcommand{\bbR}{\mathbb{R}}

\usepackage{hyperref}

\usepackage{amsmath}
    \usepackage{array}
    \usepackage{amssymb}
    \usepackage{amsfonts}
    \usepackage{bm}
    \usepackage{xspace}
    \usepackage{float}
    \usepackage{graphicx}
    \usepackage{hyperref}
    \usepackage{enumitem}
    \usepackage{color}
    \usepackage{comment}
\usepackage{appendix}
    \usepackage{mathtools}
    \usepackage{bm}
    
    \usepackage[utf8]{inputenc}
    \usepackage{balance}
    \usepackage{csquotes}
    \usepackage{microtype}
    \usepackage{breqn}
    \usepackage{titlesec}
    
    \usepackage{arydshln}
    \usepackage{booktabs}
    \usepackage{nicefrac}
    \PassOptionsToPackage{hyphens}{url}
    \usepackage{url}

\newcommand{\commentout}[1]{}

\newcommand{\Setup}{\ensuremath{\mathsf{Setup}}\xspace}

\usepackage[nointegrals]{wasysym}

\usepackage{array,multirow,booktabs,adjustbox,graphicx}

\newcommand{\ignore}[1]{}

\iftoggle{quantum}{}{
    
}

\newcommand{\ck}{\ensuremath{\mathsf{ck}}}

\usepackage{footnote}
    \makesavenoteenv{figure}

\newcommand{\bbS}{\ensuremath{\mathbb{S}}}
\newcommand{\share}{\mathsf{Share}}

\newcommand{\epr}{\mathsf{EPR}}
\newcommand{\regM}{\mathsf{M}}
\newcommand{\regT}{\mathsf{T}}

\usepackage{physics}

\newcommand{\inreg}{\ensuremath{\mathsf{Input}}}
\newcommand{\mesreg}{\ensuremath{\mathsf{Message}}}
\newcommand{\outreg}{\ensuremath{\mathsf{Output}}}

\newcommand{\rec}{\mathsf{Reconstruct}}

\iftoggle{quantum}{}{
\usepackage{soul}

}

\iftoggle{quantum}{
    \title{MPC in the Quantum Head (or: Superposition-Secure (Quantum) Zero-Knowledge)}
}{
    \title{MPC in the Quantum Head \\ \small{(or: Superposition-Secure (Quantum) Zero-Knowledge)}}
}

\iftoggle{quantum}{
    \author{Andrea Coladangelo}
    \affiliation{University of Washington}
    \orcid{0000-0002-6773-2711}
    
    \author{Ruta Jawale}
    \affiliation{University of Illinois, Urbana-Champaign}
    \orcid{0000-0002-4886-7274}
    
    \author{Dakshita Khurana}
    \affiliation{University of Illinois, Urbana-Champaign and NTT Research}
    \orcid{0000-0001-5315-4503}
    
    \author{Giulio Malavolta}
    \affiliation{Bocconi University}
    \orcid{0009-0009-9737-0094}
    
    \author{Hendrik Waldner}
    \affiliation{CISPA}
    \orcid{0000-0002-9083-5794}

}{
    \author{Andrea Coladangelo\thanks{University of Washington} \and Ruta Jawale\thanks{University of Illinois, Urbana-Champaign} \and Dakshita Khurana\thanks{University of Illinois, Urbana-Champaign and NTT Research} \and Giulio Malavolta\thanks{Bocconi University} \and Hendrik Waldner\thanks{CISPA}}
}

\date{}

\begin{document}

\maketitle

\begin{abstract}
    The MPC-in-the-head technique (Ishai et al., STOC 2007) is a celebrated method to build zero-knowledge protocols with desirable theoretical properties and high practical efficiency. This technique has generated a large body of research and has influenced the design of real-world post-quantum cryptographic signatures. In this work, we present a generalization of the MPC-in-the-head paradigm to the quantum setting, where the MPC is running a quantum computation.

    As an application of our framework, we propose a new approach to build zero-knowledge protocols where security holds even against a verifier that can obtain a \emph{superposition} of transcripts. This notion was pioneered by Damg{\r a}rd et al., who built a zero-knowledge protocol for NP (in the common reference string model) secure against superposition attacks, by relying on perfectly hiding and unconditionally binding dual-mode commitments. Unfortunately, no such commitments are known from standard cryptographic assumptions.  In this work we revisit this problem, and present two new three-round protocols in the common reference string model: (i) A zero-knowledge argument for NP, whose security reduces to the standard learning with errors (LWE) problem. (ii) A zero-knowledge argument for QMA from the same assumption.
\end{abstract}

\newpage
\tableofcontents
\newpage
\section{Introduction}

The MPC-in-the-head paradigm \cite{IshaiKOS09} is a celebrated method to build zero-knowledge (ZK) proof systems. Loosely speaking, it allows one to bootstrap a \emph{semi-honest} multi-party computation (MPC) scheme into a ZK proof system, by letting a prover run an MPC protocol \emph{in their head} and then selectively reveal parts of the transcript to the verifier, who checks for consistency in the views that have been revealed. 
The original motivation for this technique was to build \emph{constant rate} zero-knowledge protocols, but over the years this method has found more applications in the design of concretely efficient ZK proofs~\cite{DBLP:conf/ccs/BhadauriaFHVXZ20,DBLP:conf/ccs/ChaseDGORRSZ17,DBLP:conf/uss/GiacomelliMO16} and digital signature schemes~\cite{DBLP:conf/ccs/ChaseDGORRSZ17,DBLP:conf/ccs/KatzK018}.In this work, we propose a quantum generalization of the  MPC-in-the-head paradigm and show how it can be used to construct ZK protocols with security against \emph{superposition} attacks.

Superposition security is an emerging notion in cryptography that allows the adversary to query a given oracle on a \emph{superposition} of inputs, which may leak more information than classical queries alone. This notion has been fruitfully studied in the context of random oracles~\cite{BonehDFLSZ11,DonFMS19,DonFMS22}, pseudorandom functions~\cite{Zhandry12}, message authentication codes~\cite{BonehZ13,GargYZ17}, encryption and signature schemes~\cite{BonehZ132,GagliardoniHS16}, and zero-knowledge protocols~\cite{DamgardFNS13}. 
While superposition security is interesting from a theoretical standpoint as a natural generalization of ``classical-query'' security, it is also well-motivated in many practical settings. For instance, consider a computing device that runs a program on some some hidden secret information.
A quantum adversary with access to such a device can cool it down and force quantum effects on it, allowing for the aforementioned superposition attacks. Though this may at first seem implausible, similar types of (classical) attacks have already been performed in the context of side channel leakage on cryptographic hardware, which was also previously considered implausible~\cite{HaldermanSHCPCFAF08,DBLP:conf/cardis/HutterS13,YADA17}.

In the future, various (classical) primitives might natively be connected through quantum networks, either to benefit from speed-ups or because some of the parties participating in the protocol are inherently quantum. In such a setting, for example, one party may use a quantum computer to encrypt messages for another party whose prescribed implementation is classical. One might try to enforce classical communication by requiring honest parties to measure all incoming messages. However, this shifts the security of the protocol onto an additional physical assumption: that the honest party's device really performs the intended measurement, and does not preserve or process quantum coherence in a way that a quantum adversary could exploit. Proving superposition security avoids relying on such an implementation-specific assumption. Other examples in which an attacker may be able to trick a classical device into exhibiting quantum behavior arise when a classical scheme is used as a subprotocol in a larger quantum protocol~\cite{JiL018}. A prominent example of a primitive satisfying a superposition-security notion is that of superposition-secure pseudorandom functions (qPRFs)~\cite{Zhandry12}. qPRFs and related superposition-secure symmetric-key primitives have since been used as building blocks in a range of quantum-secure constructions, including quantum-secure signatures~\cite{BonehZ13}, QCCA1-secure secret-key encryption of quantum data~\cite{AlagicJOP20}, and authenticated encryption~\cite{BhaumikBCLNSS21}.Beyond these classical cryptographic applications, they also underlie constructions of pseudorandom quantum states~\cite{BrakerskiS20}, pseudorandom unitaries~\cite{BrakerskiM24}, and quantum obfuscation~\cite{BartusekM22}. This body of work suggests that superposition security is not only a natural strengthening of classical security, but also a useful design principle for primitives that are intended to be composed within larger quantum-secure protocols. A similar setting in which one could imagine using a superposition-secure zero-knowledge protocol is in GMW-style compilers~\cite{GoldreichMW87}, where zero knowledge is used to amplify the security of semi-honest multiparty computation protocols to malicious security. In the presence of a quantum adversary, one could analogously seek such a compiler that provides security amplification even against superposition attacks. Such a compiler would naturally require the underlying zero-knowledge protocol to remain secure under superposition access. We therefore view superposition-secure zero-knowledge as a natural addition to this landscape.

\paragraph{ZK under Superposition Attacks.} In this work, we build proof systems whose zero-knowledge property holds even when the quantum verifier is allowed to issue several different challenges to the prover in superposition, and to receive the corresponding responses in superposition.
The only known proof system in this setting, proposed by Damg{\r a}rd et al.~\cite{DamgardFNS13}, is a three-round protocol based on a perfectly hiding and binding dual-mode commitment. Unfortunately, such commitments are not known to exist from standard cryptographic assumptions. Furthermore, their proof technique does not appear to easily generalize to prove QMA statements, the natural quantum analogue of NP.
Our work aims to address these drawbacks.

As already observed by Damg{\r a}rd et al.~\cite{DamgardFNS13}, any non-interactive zero-knowledge (NIZK) proof is \emph{trivially} superposition secure, since the verifier does not make any ``queries'' to the prover. 
However, to the best of our knowledge, NIZKs for QMA are not known to exist under any standard cryptographic assumptions. The main candidate NIZK for QMA that we are aware of~\cite{BartusekM22} is in an oracle model and relies on strong heuristic assumptions about classical obfuscation. Another NIZK construction~\cite{AlagicCGH20}, in the quantum random oracle model, requires designated verifiers which results in interaction between the prover and the verifier during the setup procedure. To adequately adapt the model of superposition security to this setting, we would need to prove security against an adversary that can ask superposition queries in the setup as well. We leave this as future work and focus on the setting where the setup only outputs public information.

Furthermore, even in the simpler setting of ZK for NP, our three-round protocols are essentially specialized $\Sigma$-protocols; these are significantly more lightweight than (post-quantum) NIZKs.

\paragraph{Our Contributions.}
We build ZK with superposition security for languages in QMA, based on the learning with errors (LWE) assumption.
The main building block needed for this result, and the main technical contribution of this work, is the first instantiation of the MPC-in-the-head paradigm~\cite{IshaiKOS09} for quantum functionalities.

\begin{theorem}[Informal]
Assuming the post-quantum security of LWE, there exist three-round $\Sigma$-protocols for QMA in the CRS model satisfying zero-knowledge against malicious verifiers that make superposition queries to a prover.	\end{theorem}

As a warmup, we consider the (weaker) setting of superposition-secure ZK for languages in {\em NP}. Here, we obtain $\Sigma$-protocols satisfying superposition security against malicious verifiers based on {\em any} dual-mode commitment scheme, allowing for instantiations based on the learning with errors (LWE) assumption. This improves prior work~\cite{DamgardFNS13} which relied on perfectly binding dual-mode commitments, that were not known to exist based on standard assumptions. This leads to the following (informal) theorem.
\begin{theorem}[Informal] Assuming the existence of any post-quantum dual-mode commitments, there exist three-round $\Sigma$-protocols for NP in the CRS model satisfying zero-knowledge against malicious verifiers that make superposition queries to a prover.
\end{theorem}
Note that post-quantum dual-mode commitments exist from various assumptions, including LWE \cite{PeikertVW08}, which implies that our results can be based on the standard post-quantum hardness of the LWE problem.

\subsection{Our Techniques}

In the following, we provide a somewhat informal and brief overview of our techniques, and we refer the reader to the subsequent sections for more details.

\subsubsection{Superposition-Secure Zero-Knowledge for NP}
\label{sec:tech-overview-np}

Before presenting our protocol, it is useful to recap the three-round protocol of~\cite{DamgardFNS13}, based on the MPC-in-the-head paradigm. Their protocol is the following: 
\begin{itemize}
\item Let $x$ be the instance of the problem. To compute its first message, the prover, on input a witness $w$, runs, in its head, an MPC protocol that computes the NP relation $\mathcal{R}(x, m_1 \oplus \cdots \oplus m_n)$ where $m_i$ is the input of the $i$-th party and $m_1 \oplus \cdots \oplus m_n = w$ (in other words, one can think of $m_1, \dots, m_n$ as shares of an $n$-out-of-$n$ secret sharing of the witness). Then, the prover generates commitments to the views of the parties of the previously executed protocol and sends them over to the verifier as the first message. 
\item The verifier asks the prover to open a uniformly sampled subset of these commitments. 
\item The prover responds with the corresponding openings. If the openings are valid, the verifier checks that the views are correct (by checking that all local operations are executed correctly and that all messages are passed consistently across parties), and that the output of the prover's MPC protocol was ``accept''. If so, the verifier outputs ``accept''.
\end{itemize}

The soundness of this protocol, in the presence of quantum adversaries, follows by appealing to the (perfect) robustness of the MPC (similarly as in the classical case)\footnote{An MPC protocol is $t$-robust if any set of corrupted parties of size at most $t$ cannot make an honest party output the incorrect output of the circuit being computed.}. On the other hand, zero-knowledge needs to be proven against a malicious verifier that asks the prover for a superposition of openings. This is shown in two steps:\begin{itemize}
    \item First, since the underlying MPC protocol achieves perfect privacy, the messages of the prover can be viewed as a secret sharing of the witness, and therefore remains hidden as long as the verifier only receives openings of a small enough subset of views.
    \item Second, and this is the key novelty in the argument of Damg{\r a}rd et al.~\cite{DamgardFNS13}, one invokes the following lemma.
    \begin{lemma}[\cite{DamgardFNS13}, informal]
    \label{lem:dfns}
        A $t$-out-of-$n$ perfectly secure secret sharing scheme remains perfectly secure even if the distinguisher queries the shares in superposition, as long as less than $2t$ shares are queried.
    \end{lemma}
\end{itemize}
We highlight that this approach crucially relies on the fact that the 
commitment scheme being used is \emph{perfectly} hiding. Otherwise, one cannot directly argue that the opened views of the MPC, together with the unopened commitments, are a perfectly secure secret sharing scheme. For instance, if the secret sharing scheme is only computationally hiding, then an unbounded adversary can simply learn the shares that the prover did not open. Quite surprisingly though, even for a \emph{statistically} secure secret sharing scheme 
it is not easy to argue that the resulting secret sharing scheme 
resists superposition attacks! Indeed, as far as we know, it is an interesting open question to determine whether Lemma \ref{lem:dfns} holds if we replace ``perfect'' with ``statistical'' security.\footnote{Morally, Lemma \ref{lem:dfns} is proven using a similar idea as a subsequent result by Zhandry \cite{zhandry2015secure} showing that $q$ quantum queries to a uniformly random oracle can be simulated using a $2q$-wise family of hash functions. There, the result is also not known to extend to approximate $2q$-wise hash functions. This is because, in the statistical case, the security loss is the sum of exponentially many (exponentially small) losses. These are all zero in the perfect case. An avid reader might think that subexponential security might help to overcome this issue but this still does leave open the question of the superposition security for statistical secure secret sharing schemes.}

\paragraph{Our Technique.} In this work, we overcome the aforementioned issue via a new proof strategy. We rely on {\em dual-mode commitments}, where a common reference string is sampled in one of two indistinguishable modes. The commitment is statistically binding in one mode and statistically (but not perfectly) hiding in the other.  In order to establish the zero knowledge property, we rely (only) on the statistically hiding mode of the commitment.
In the proof, we introduce an additional hybrid where all of the commitments that the prover sends are commitments to $0$ (as opposed to the honestly generated MPC views). Since these commitments are statistically hiding, and thus only computationally binding, they can still be {\em opened} by the prover to the same honest MPC views as before (with a small statistical security loss). Of course, the latter cannot be done efficiently, but this is fine for the purposes of the reduction in this proof, since an efficient quantum verifier that distinguishes the two hybrids still implies an (inefficient) adversary that breaks the statistical hiding property.\footnote{To provide answers to quantum queries here, we apply the same operation (modeled as a unitary), controlled on the received quantum state.} Once we have made this change, we observe that the commitments to 0 generated by the prover, together with (a small enough) subset of openings to honest MPC views, correspond to (a small enough number of) shares of a \emph{perfectly} secure secret sharing scheme. From here on, we can conclude the security proof by applying similar arguments as in~\cite{DamgardFNS13}. The formal description and analysis of this protocol can be found in~\cref{sec:zk-np}.

\subsubsection{MPC in the Quantum Head}

Upgrading the above protocol to a superposition-secure zero-knowledge protocol for QMA presents the following challenge: a QMA verification circuit is quantum, and the witness is a quantum state, so we need to consider a multiparty \emph{quantum} computation protocol (MPQC) rather than an MPC protocol. However, in adapting the MPC-in-the-head paradigm to an MPQC protocol, we run into the following fundamental obstacle: since an MPQC protocol involves quantum inputs computation and communication, the views of the parties in the protocol are no longer classical, and there is no well-defined notion of a \emph{transcript} that the verifier can check for consistency. In a bit more detail, in each round of an MPQC protocol, each party receives quantum messages from another party and operates on them using its private quantum state. Therefore, at the end of the protocol, a party is not in possession of a full transcript of the execution but only of its final state! This makes it harder to verify the correctness of the computation. We resolve this obstacle with the following idea: \begin{itemize}
\item We view the global MPQC protocol, including the exchange of messages, as a quantum circuit (where message passing is substituted by SWAP gates). We then apply Kitaev's circuit-to-Hamiltonian reduction \cite{KSV02} to this quantum circuit to obtain a local Hamiltonian (recall that the local terms of this Hamiltonian serve to enforce the correct execution of the computation). The ground state of this Hamiltonian, also referred to as the \emph{history state}, has low-energy \emph{if and only if the MPQC protocol accepts} (on a well-formed input), i.e.\ if and only if the QMA verification has a valid witness.
\item Similarly as in \cite{broadbent2020zero}, the prover commits to the \emph{history state} of the Hamiltonian (this commitment is done qubit-by-qubit, so that an arbitrary subset of qubits can be opened while the others stay hidden). The verifier then asks the prover to open the qubits corresponding to one of the Hamiltonian terms. Upon receiving the opening, the verifier can measure the Hamiltonian term, and check that the energy is below a certain threshold. This check, averaged over all of the Hamiltonian terms, ensures that the history state corresponds to a correctly executed MPQC protocol (with an accepting output), and thus ensures soundness.
\item The ZK property follows from the following crucial observation: opening only the qubits corresponding to any one of the local checks of the Hamiltonian does not reveal any more information about the witness than what could be learnt by controlling a small subset of parties in the MPQC protocol. Since the MPQC protocol (similarly to the classical MPC-in-the-head setting) is running the QMA verification circuit on a \emph{secret-sharing} of the witness, this ensures that no information about the witness is revealed.
\end{itemize}

We call this approach \emph{MPC in the quantum head}. This can be seen as a generalization and abstraction of the approach taken by Broadbent and Grilo in \cite{broadbent2022qma}. In more detail, the locally simulatable proofs they introduce, relying on the circuit-to-Hamiltonian reduction and the local Hamiltonian problem in the same way as we do, realizes similar properties to the properties we rely on in this work, i.e., allowing for verification while only revealing parts of the witness. In order to instantiate our approach, we identify a suitable notion of MPQC, and show that the MPQC protocol of Crepeau et al.~\cite{CGS02}, with an adapted security analysis (described in~\cref{sec:prelim-mpqc}), suffices for our purpose.

This generalization of the approach of Broadbent and Grilo~\cite{broadbent2022qma} using MPQC~\cite{CGS02} is the main conceptual contribution of this work. We believe that this technique can be extended to other commit-and-open~\cite{EPRINT:Chailloux19,C:DFLSS09,C:DFMS22} and cut-and-choose protocols~\cite{DEC:CLLW22,ARXIV:KasMusWal17,ARXIV:WiePap26,CiC:WCKPK26}, as well as to other protocols requiring partial verification of quantum computations. Besides the applications to security against superposition attacks, we envision that such a technique could be used to show QMA-hardness of problems beyond CLDM~\cite{broadbent2022qma}. We leave this exploration as a fascinating open research direction.

We provide a bit more detail about our MPC-in-the-quantum-head approach. In our MPQC protocol (which is meant to be executed ``in the head''), the QMA witness is secret-shared across the parties in the following way: the witness $\ket{w}$ is encrypted using a quantum one-time pad $\ket{v}=X^{a}Z^{b}\ket{w}$ and shares $(a_i,b_i)$ of the key are distributed among the first $n-1$ parties. The $n^{th}$ party (which we shall think of as a dummy party) does not receive a key share but the quantum one-time pad encryption $\ket{v}$ instead. The MPQC executes the circuit that 
\begin{itemize}
\item[(i)] Recombines all the key shares $(a_i,b_i)$ of the first $n-1$ parties to obtain the key $(a,b)$, and uses it to decrypt the ciphertext $\ket{v}$, which is provided by the $n^{th}$ party, to obtain the witness $\ket{w}$. 
\item[(ii)] Runs the QMA verification circuit on the recovered witness $\ket{w}$. 
\end{itemize}

As outlined above, the prover in our ZK protocol commits to a \emph{history state} of the Hamiltonian corresponding to the ``global'' MPQC circuit. This commitment is also done using a quantum one-time pad: to commit to the history state $\ket{\psi_{hist}}$, apply a uniformly random one-time pad to obtain $\ket{\phi}=X^aZ^b\ket{\psi}$, and then use the dual-mode commitment to commit to the (classical) keys $a,b$. This results in a commitment scheme that is \emph{dual-mode}: in one mode it is statistically hiding and computationally binding, and in the other it is perfectly binding and computationally hiding. Furthermore, the two modes are computationally indistinguishable.

The proof of the ZK property of this protocol uses a similar technique as the one we described in Section \ref{sec:tech-overview-np} to obtain a superposition-secure ZK protocol for NP. The key difference is that now, instead of introducing a hybrid where the prover commits to 0's, like we did earlier, we introduce a hybrid where all of the commitments are replaced with commitments to half of an EPR pair. This allows an unbounded prover to open each commitment to an arbitrary quantum state $\ket{\psi}$ as follows: teleport $\ket{\psi}$ into the committed register, and then inefficiently open the (computationally binding) commitments of the one-time pad keys to the teleportation errors. 

An interesting difference compared to the classical setting is that our MPQC-in-the-head framework does \emph{not} require the underlying MPC to be perfectly correct, nor robust, as was the case for~\cite{IshaiKOS09}. This is due to the fact that we can ``force'' parties in the MPQC to sample their random coins honestly (for instance, by measuring a $\ket{+}$ state), whereas no such procedure exists classically.

\section{Preliminaries}

In this section, we introduce all the necessary preliminaries for our protocols. We start with the local Hamiltonian problem.

\subsection{Local Hamiltonian Problem is $\QMA$-complete}

\begin{definition}[\protect $\QMA$~\cite{KR03}]
    \label{def:qma}
    Let $\cB$ be the Hilbert space of a qubit.
    Fix $\epsilon(\cdot)$ such that $2^{-\Omega(\cdot)} \le \epsilon(\cdot) \le \frac{1}{3}$. 
Then, a promise problem $\cL=(\cL_{yes}, \cL_{no}) \in \QMA$ if there exists a quantum polynomial-size family of circuits $\cM = \{\cM_n\}_{n \in \bbN}$ and a polynomial $p(\cdot)$ such that:
    \begin{itemize}
        \item For all $x \in \cL_{yes}$, there exists $\ket{\xi} \in \cB^{\tensor p(|x|)}$ such that $\Pr[\cM_{|x|}(\ket{x}, \ket{\xi}) = 1] \ge 1 - \epsilon(|x|)$.
        \item For all $x \in \cL_{no}$, and for all $\ket{\xi} \in \cB^{\tensor p(|x|)}$ it holds that $\Pr[\cM_{|x|}(\ket{x}, \ket{\xi}) = 1] \le \epsilon(|x|)$.
    \end{itemize}
\end{definition}

\begin{definition}[$2$-local Hamiltonian problem]
    \label[definition]{def:2locHam}
The $2$-local Hamiltonian problem with functions $a, b$ where $b(n) > a(n)$ for all $n \in \bbN$
 is the promise problem $\cL = (\cL_{yes}, \cL_{no})$ defined as follows. An instance is a Hermitian operator on some number $n$ of qubits, taking the following form:
    \begin{equation*}
        H = \sum_{S} d_S S
    \end{equation*}
    where each $d_S$ is a real number and $S$ is a ``$2$-local'' operator, i.e., it acts non-trivially only on $2$ out of the $n$ qubits.\begin{itemize}
        \item $H \in \cL_{yes}$ if the smallest eigenvalue of $H$ is at most $a(n)$.
        \item $H \in \cL_{no}$ if the smallest eigenvalue of $H$ is at least $b(n)$.
    \end{itemize}
    \begin{theorem}[$2$-local Hamiltonian is QMA-complete \cite{KKR06}]
        The 2-local Hamiltonian problem with functions $a,b$ is QMA-complete if $b(n) - a(n) > 1/\poly(n)$.
    \end{theorem}

\end{definition}

\subsection{Circuit-to-Hamiltionian Reduction}
\label{sec:history}

We recall the notion of a Circuit-to-Hamiltonian reduction, and of a history state.

\begin{lemma}[Lemma 3 of~\cite{KKR06}]
\label{lem:circuit-to-hamiltonian}
There exists an efficiently computable map from quantum circuits $\cC$ to 2-local Hamiltonians $H_{\cC}$, which we refer to as a ``Circuit-to-Hamiltonian reduction'', that satisfies the following. Let $\epsilon>0$.
\begin{enumerate}[label=(\roman*)]
\item If there exists a state $\ket{\nu}$ such that $\cC$ outputs $1$ on input $\ket{\nu, 0}$ with probability at least $1-\epsilon$, then $H_{\cC}$ has an eigenvalue smaller than $\epsilon$. Moreover, suppose that $\cC = U_T\dots U_1$ for some 2-qubit unitaries $U_1,\ldots, U_T$. Then, the state 
 \[
        \ket{\phi_{hist}}=\frac{1}{\sqrt{T+1}}\sum_{t=1}^TU_t\dots U_1\ket{\nu,0}\otimes\ket{t}
    \]
is the ground state of $H_{\cC}$ (i.e.\ the eigenvector with the lowest eigenvalue). We refer to this state as the ``history state''.

\item  if, for all $\ket{\nu}$, $\cC$ outputs $1$ on input $\ket{\nu, 0}$ with proability at most $\epsilon$, then all eigenvalues of $H_{\cC}$ are larger than $\frac{1}{2}-\epsilon$. \end{enumerate}
\end{lemma}

By \cref{def:2locHam}, the Hamiltonian $H$ can be written as $H = \sum_S d_S S$ where $d_S \in \bbR$ and $S$ is a tensor product of Pauli operators where only two operators are $Z$ or $X$, and others are $\identity$.

In order to perform a verification, we need to convert $H$ to a measurement and discuss the new thresholds.
We can rescale $H$ as 
\begin{equation*}
    H' = \sum_S p_S \left(\frac{\identity + \text{sign}(d_S) S}{2}\right)
\end{equation*} 
where $p_S = \frac{|d_S|}{\sum_S |d_S|} \in [0,1]$.

We now outline a verification scheme for a purported history state $\rho$. The verification that $\rho$ is a ground state of $H'$ involves repeating the following check:
\begin{itemize}
    \item Sample a term $S$ with probability $p_S$.
    \item Apply the measurement $\{M_1 = \frac{\identity + S}{2}, M_{-1} = \frac{\identity - S}{2}\}$ to get outcome $e \in \{1,-1\}$. Note that this measurement involves only the two qubits that $S$ acts non-trivially on. \item Output $\accept$ iff $e = -\text{sign}(d_S)$.
\end{itemize}
On input a state $\rho$, the probability that the protocol accepts is $1 - \mathsf{Tr}[H'\rho] = \frac12 - \frac{\mathsf{Tr}[H \rho]}{2\sum_S |d_S|}$.

\subsection{Commitment Schemes and Quantum One-Time Pad}

\begin{definition}[Commitment Scheme] 
\label[definition]{def:com}
$\cC = (\Gen, \Com)$ is a commitment scheme if it has the following syntax:
\begin{itemize}
    \item $\ck \gets \Gen(1^{\lambda})$: The classical polynomial-time  key generation algorithm on input security parameter $1^{\lambda}$ outputs public commitment key $\ck$.
    \item $c \gets \Com(\ck, m; r)$: The classical polynomial-time commitment algorithm on input commitment key $\ck$, message $m$ and randomness $r$ outputs commitment $c$.
\end{itemize}
$\cC$ has the following properties:
\begin{itemize}
    \item \textbf{Perfect Binding}: 
    For all messages $m$, $m'$ such that $m \ne m'$, all randomness $r$, $r'$, and all $\lambda \in \bbN$,
    \begin{equation*}
       \Pr_{\ck \gets \Gen(1^\lambda)}[\Com(\ck, m; r) = \Com(\ck, m'; r')] = 0.
    \end{equation*}

    \item \textbf{Computational Hiding}: 
    For all quantum polynomial-size circuits $\cD = \{\cD_\lambda\}_{\lambda \in \bbN}$, there exists a negligible function $\negl(\cdot)$ such that
for all messages $m$ and $m'$, and all $\lambda \in \bbN$,
    \begin{equation*}
        \left\vert \Pr_{\substack{\ck \gets \Gen(1^{\lambda}) \\ c \gets \Com(\ck, m)}}[\cD_\lambda(\ck, c) = 1] - \Pr_{\substack{\ck \gets \Gen(1^{\lambda}) \\ c' \gets \Com(\ck, m')}}[\cD_\lambda(\ck, c') = 1]\right\vert \le \negl (\lambda).
    \end{equation*}

\end{itemize}
\end{definition}

As an instantiation for the described commitments, we can rely on the work of~\cite{LombardiS19}, in which the authors present non-interactive commitments based on the LWE assumption.

\begin{definition}[Dual-mode Commitment]
    \label[definition]{def:dm-com}
    $\cC = (\Gen_B, \Gen_H, \Com)$ is a dual-mode commitment scheme if it has the following syntax:
\begin{itemize}
    \item $\ck_B \gets \Gen_B(1^{\lambda})$: In the binding mode, the classical polynomial-time key generation algorithm on input security parameter $1^{\lambda}$ outputs public commitment key $\ck_B$.
    
    \item $\ck_H \gets \Gen_H(1^{\lambda})$: In the hiding mode, the classical polynomial-time key generation algorithm on input security parameter $1^{\lambda}$ outputs public commitment key $\ck_H$.
    
    \item $c \gets \Com(\ck, m; r)$: The classical polynomial-time commitment algorithm on input commitment key $\ck$ (generated either in binding or hiding mode), message $m$ and randomness $r$ outputs commitment $c$.
\end{itemize}
$\cC$ has the following properties: \begin{itemize}

    \item \textbf{Indistinguishability of Modes:}
For all quantum polynomial-size circuits $\cD = \{\cD_\lambda\}_{\lambda \in \bbN}$, there exists a negligible function $\negl(\cdot)$ such that for all $\lambda \in \bbN$,
    \begin{equation*}
        \left\vert \Pr_{\substack{\ck_H \gets \Gen_H(1^{\lambda})}}[\cD_\lambda(\ck_H) = 1] - \Pr_{\substack{\ck_B \gets \Gen_B(1^{\lambda})}}[\cD_\lambda(\ck_B) = 1]\right\vert \le \negl(\lambda).
    \end{equation*}

    \item \textbf{Perfect Binding}: 
    For all messages $m$, $m'$ such that $m \ne m'$, all randomness $r$, $r'$, and all $\lambda \in \bbN$,
    \begin{equation*}
       \Pr_{\ck_B \gets \Gen_B(1^\lambda)}[\Com(\ck_B, m; r) = \Com(\ck_B, m'; r')] = 0.
    \end{equation*}
    
    \item \textbf{Statistical Hiding}: For all quantum unbounded-size circuits $\cD = \{\cD_\lambda\}_{\lambda \in \bbN}$, there exists a negligible function $\negl(\cdot)$ such that for all messages $m$ and $m'$, and all $\lambda \in \bbN$,
    \begin{equation*}
        \left\vert \Pr_{\substack{\ck_H \gets \Gen_H(1^{\lambda}) \\ c \gets \Com(\ck_H, m)}}[\cD_\lambda(\ck_H, c) = 1] - \Pr_{\substack{\ck_H \gets \Gen_H(1^{\lambda}) \\ c' \gets \Com(\ck_H, m')}}[\cD_\lambda(\ck_H, c') = 1]\right\vert \le \negl(\lambda).
    \end{equation*}

\end{itemize}
\end{definition}

To instantiate the above described dual-mode commitment scheme, we rely on the work of Peikert, Vaikuntanathan and Waters~\cite{PeikertVW08} where the authors construct a LWE-based dual-mode cryptosystem, which can be cast as a commitment scheme.

\begin{definition}[Quantum One-time Pad (Quantum OTP)]
   Let $m \in \bbN$ and $\ket{\varphi}$ an $m$-qubit state be given. Let $a$ and $b$ be sampled uniformly at random from $\zo^m$. Let $\ket{\psi} = X^a Z^b \ket{\varphi}$. We say that $\ket{\varphi}$ is the Quantum OTP of $\ket{\psi}$ with keys $(a, b)$.
\end{definition}

\begin{lemma}
    Let $m \in \bbN$ and $\ket{\varphi}$ an $m$-qubit state be given. We have that 
    \begin{equation*}
        \frac{1}{2^{2m}}\sum_{a, b \in \zo^m} X^a Z^b \ket{\varphi}\bra{\varphi}Z^b X^a = \frac{1}{2^m} \identity_m.
    \end{equation*}
\end{lemma}

\subsection{Zero-Knowledge}

In this section, we define classical zero-knowledge notions as well as their quantum analogues in parenthesis.

\begin{definition}[Zero-Knowledge Argument (for Promise Problem)]
\label[definition]{def:zk}
Let $\NP$ relation $\cR$ with corresponding language $\cL$ be given such that they can be indexed by a security parameter $\lambda \in \bbN$.
(Let $\QMA$ promise problem $\cL = (\cL_{yes}, \cL_{no})$ be given with a corresponding deciding quantum polynomial-size circuit $\cM$.) 

$\Pi = (\Setup, \sP, \sV)$ is a post-quantum (quantum) zero-knowledge argument for $\NP$ (for $\QMA$) in the CRS model if it has the following syntax and properties.

\paragraph{Syntax.}
The input $1^\lambda$ is left out when it is clear from context.
\begin{itemize}
    \item $\Setup$ is a probabilistic polynomial-size circuit that on input $1^\lambda$ outputs a common reference string $\crs$.
    \item $\sP$ is an interactive probabilistic (quantum) polynomial-size circuit that takes as input a common reference string $\crs$ and a pair $(x, w) \in \cR_\lambda$ (where $x \in \cL_{yes}$ with $|x| = \lambda$, and  $\ket{w}$ is a valid witness).
    \item $\sV$ is an interactive probabilistic (quantum) polynomial-size circuit $\sV$ that takes as input a common reference string $\crs$, an instance $x$, and outputs $0$ or $1$.
\end{itemize}

In the following, we use the notation $\langle \sP(w), \sV \rangle(\crs, x)$ to denote $\sV$'s output given an interaction with $\sP$ where $(\crs, x)$ are shared inputs and $w$ is a private input to $\sP$.

\paragraph{Properties.}
\begin{itemize}
    \item {\bf Completeness:}
    There exists a negligible function $\negl(\cdot)$ such that every $\lambda \in \bbN$ and every $(x, w) \in \cR_\lambda$ (every $x \in \cL_{yes}$ with a valid witness $\ket{w}$ and where $|x| = \lambda$),
\begin{align*}
        &\Pr_{\substack{\crs \gets \Setup(1^\lambda)}}[\langle \sP(w), \sV \rangle(\crs, x) = 1] = 1\\
        &\left(\Pr_{\substack{\crs \gets \Setup(1^\lambda)}}[\langle \sP(\ket{w}), \sV \rangle(\crs, x) = 1] \ge 1 - \negl(\lambda)\right).
    \end{align*}

    \item {\bf Computational Soundness:}
    For every polynomial-size family of quantum circuits 
    $\cA = \{\cA_\lambda\}_{\lambda \in \bbN}$, there exists a negligible function $\negl(\cdot)$ such that for every $\lambda \in \bbN$, and every $x \not\in \cL_\lambda$ (every $x \in \cL_{no}$ where $|x| = \lambda$),
    \begin{align*}
        &\Pr_{\substack{\crs \gets \Setup(1^\lambda)}}[\langle \cA_{\lambda}, \sV \rangle(x, \crs) = 1]  \le \negl(\lambda).
\end{align*}

    \item {\bf Computational Zero-Knowledge:}
    There exists a probabilistic (quantum) polynomial-size circuit $\Sim = (\Sim_0, \Sim_1)$ such that for all polynomial-size quantum circuit $\cA$, there exists a negligible function $\negl(\cdot)$ for all $\lambda \in \bbN$, and all $(x, w) \in \cR_\lambda$ (all $x \in \cL_{yes}$ with a valid witness $\ket{w}$ and where $|x| = \lambda$), 
\begin{align*}
        &\left\vert \Pr_{\substack{\crs \gets \Setup(1^\lambda)}}[\langle \sP(w), \cA \rangle(\crs, x) = 1] - \Pr_{\substack{(\crs, \td) \gets \Sim_0(1^\lambda)}}[\langle \Sim_1(\td), \cA \rangle(\crs, x) = 1]\right\vert \le \negl(\lambda)\\
        &\left(\left\vert \Pr_{\substack{\crs \gets \Setup(1^\lambda)}}[\langle \sP(\ket{w}), \cA \rangle(\crs, x) = 1] - \Pr_{\substack{(\crs, \td) \gets \Sim_0(1^\lambda)}}[\langle \Sim_1(\td), \cA \rangle(\crs, x) = 1]\right\vert \le \negl(\lambda)\right).
    \end{align*}
\end{itemize}
\end{definition}
Next, we define the notion of superposition security for zero-knowledge arguments. For notational convenience, we only define it for the special case of three-round protocols, although a more general definition can be derived by extending the syntax accordingly.
\begin{definition}[Superposition-Secure Zero-Knowledge Argument]
A protocol $\Pi = (\sP, \sV)$ is a post-quantum (quantum) superposition-secure zero-knowledge argument if it is a post-quantum (quantum) zero-knowledge argument (\cref{def:zk}) where the zero-knowledge adversary is additionally given superposition-query access, and the simulator is a quantum polynomial-size circuit. Specifically, let $\Sim^*_2(c) \to r_c$ be next-message function of the simulator that, on input a challenge $c \in \mathcal{C}$, returns a response $r_c$, and let
    \[
    U_{\Sim_2} \ket{c} \ket{y} \to \ket{c} \ket{y \oplus r_c}
    \]
    be its unitary implementation. Then we require that zero-knowledge holds even if the distinguisher can query $U_{\Sim_2}$ once.
\end{definition}

\subsection{Secret Sharing Scheme}
\label{sec:prelim-sss}

\begin{definition}[$t$-out-of-$n$ Secret Sharing Scheme]
    \label[definition]{def:sss}
    $S = (\mathsf{Share}, \mathsf{Reconstruct})$ is a $t$-out-of-$n$ secret sharing scheme over the set of all possible secrets $\bbS$ and randomness $\cR$ if it has the following syntax and properties.

    \paragraph{Syntax.}
    \begin{itemize}
        \item $(s_1, \ldots, s_n) \gets \mathsf{Share}(m; r)$: The classical polynomial-time share algorithm that on any input secret $s \in \bbS$ and randomness $r \in \cR$ outputs $n$ shares $(s_1, \ldots, s_n)$.
        \item $s \gets \mathsf{Reconstruct}(s_1, \ldots, s_n)$: The classical polynomial-time reconstruct algorithm that on any input of $n$ shares $(s_1, \ldots, s_n)$ will output a secret $s \in \bbS$.
    \end{itemize}

    \paragraph{Properties.}
    \begin{itemize}
        \item \textbf{Perfect $t$-Correctness:} For all $s \in \bbS$, for all $I \subseteq [n]$ where $|I| = t$,
        \begin{equation*}
            \Pr_{(s_1, \ldots, s_n) \gets \mathsf{Share}(s)}[\mathsf{Reconstruct}((s_i)_{i \in I}) = s] = 1.
        \end{equation*}

        \item \textbf{Perfect $(t-1)$-Security:} For all $s, s' \in \bbS$, for all $I \subseteq [n]$ where $|I| \le t - 1$,
        \begin{equation*}
            \{(s_i)_{i \in I}\}_{(s_1, \ldots, s_n) \gets \mathsf{Share}(s)} = \{(s_i')_{i \in I}\}_{(s_1', \ldots, s_n') \gets \mathsf{Share}(s')}.
        \end{equation*} \end{itemize}
\end{definition}

\subsection{Multiparty (Quantum) Computation}
\label{sec:prelim-mpqc}

\begin{definition}[Local consistency~\cite{IshaiKOS09}]
    We say that a pair of views $V_i$, $V_j$ are consistent (with respect to the protocol $\Pi$ and some public input $x$) if the outgoing messages implicit in $V_i$ are identical to the incoming messages reported in $V_j$ and vice versa.
\end{definition}

\begin{lemma}[Local vs.\ global consistency~\cite{IshaiKOS09}]
    Let $\Pi$ be an $n$-party protocol as above and $x$ be a public input.
    Let $V_1, \ldots, V_n$ be an n-tuple of (possibly incorrect) views. Then all pairs of views $V_i$, $V_j$ are consistent with respect to $\Pi$ and $x$ if and only if there exists an honest execution of $\Pi$ with public input $x$ (and some choice of private inputs $w_i$ and random inputs $r_i$) in which $V_i$ is the view of $P_i$ for every $1 \le i \le n$.
\end{lemma}

\begin{definition}[Secure Multiparty Computation (MPC)~\cite{IshaiKOS09}]
    \label[definition]{def:mpc}
    $\Pi$ is MPC protocol with if it has the following syntax and properties.

    \paragraph{Syntax:}
    \begin{itemize}
        \item There are $n$ parties $P_1, \ldots, P_n$ which all share a public input $x$. Each party, for $i \in [n]$, $P_i$ holds a private input $w_i$ and randomness $r_i$.
        \item It securely realizes an $n$-party functionality $f$, where $f$ maps the inputs $(x, w_1, \ldots, w_n)$ to an $n$-tuple of outputs.
        \item It is specified via its next message function. That is, $\Pi(1^\lambda, i, x, w_i, r_i, (m_1, \ldots, m_j))$ returns the set of $n$ messages (and, possibly, a broadcast message) sent by $P_i$ in Round $j + 1$ given security parameter $\lambda$, the public input $x$, its local input $w_i$, its random input $r_i$, and the messages $m_1, \ldots, m_j$ it received in the first $j$ rounds. 
\item The output of $\Pi$ may also indicate that the protocol should terminate, in which case $\Pi$ returns the local output of $P_i$ for its input $i \in [n]$. 
        \item The view of $P_i$, for $i \in [n]$, denoted by $V_i$, includes $(w_i, r_i)$ and the messages received by $P_i$ during the execution of $\Pi$. 
\end{itemize}
    \paragraph{Properties:}
    \begin{itemize}
        \item \textbf{Perfect Correctness}: For all inputs $x, w_1, \ldots, w_n$, the probability that the output of some player is different from the output of $f$ is $0$, where the probability is over the independent choices of the random inputs $r_1, \ldots, r_n$.

        \item \textbf{Perfect $t$-Robustness:} 
        We say that $\Pi$ realizes $f$ with perfect $t$-robustness if it is perfectly correct in the presence of a semi-honest adversary, and furthermore for any computationally unbounded malicious adversary corrupting a set $T$ of at most $t$ players, and for any inputs $(x, w_1, \ldots, w_n)$, the following robustness property holds. If there is no $(w'_1, \ldots, w'_n)$ such that $f(x, w'_1, \ldots, w'_n) = 1$, then the probability that some uncorrupted player outputs $1$ in an execution of $\Pi$ in which the inputs of the honest players are consistent with $(x, w_1, \ldots, w_n)$ is $0$.

        \item \textbf{Perfect $t$-Privacy:} Let $1 \le t \le n$. We say that $\Pi$ realizes $f$ with perfect $t$-privacy if there is a probabilistic polynomial-size circuit simulator SIM such that for any inputs $x, w_1, \ldots, w_n$ and every set of corrupted players $T \subseteq [n]$, where $|T| \le t$, the joint view $\mathsf{View}_T(x, w_1, \ldots, w_n)$ of players in $T$ is distributed identically to $\mathsf{SIM}(T, x, (w_i)_{i \in T}, f_T (x, w_1, \ldots, w_n))$. 
        
\end{itemize}
\end{definition}

\begin{theorem}[MPC~\cite{Ben-OrGW88}]
    Let $n \in \bbN,t < n/3$ and $t' < n/2$, then there exists an $n$-party, $t$-robust, and perfectly $t'$-private MPC protocol (as in \cref{def:mpc}).
\end{theorem}
\begin{proof}[Proof (Sketch)]
    To prove this theorem, we rely on the protocol of Ben-Or, Goldwasser and Wigderson~\cite{Ben-OrGW88}. We start by analysing the robustness of the protocol. In~\cite[Theorem 3]{Ben-OrGW88}, the authors show that for $t<n/3$ their protocol is $t$-resilient. Since the notion of resilience directly corresponds to the notion of robustness, the first half of the theorem follows.

    To argue the privacy case, we rely on~\cite[Theorem 1]{Ben-OrGW88}, where the authors show that their protocol achieves $t'$-privacy for $t'<n/2$ against semi-honest adversaries. Since their protocol does not require any computational assumptions, perfect privacy directly follows.
\end{proof}

We now introduce a definition for a multi-party computation in the quantum setting. This definition, unlike our classical definition, is written in the real and ideal world paradigm.

\begin{definition}[MPQC Syntax]
    \label[definition]{def:mpqc-syntax}
    For each round of a MPQC protocol, each party computes a local unitary on its input and exchange message registers with all other parties. We formalize this for an $n$-party, $K$-round MPQC protocol as follows: \begin{itemize}
        \item \textbf{Registers}: The $i$th party computes on one input register $\inreg_i$, $Kn$ message registers for each party and each round of the protocol $\{\mesreg_{i, j, k}\}_{j \in [n\setminus \{i\}], k \in [K]}$, and one output register $\outreg_i$.
        \item \textbf{Rounds}: During the $k$th round, a unitary $U_{i,k}$ is applied to the $i$th parties' registers for all $i \in [n]$, and a swap operation between $\mesreg_{i, j, k}$ and $\mesreg_{j, i, k}$ for all pairs of parties $i,j \in [n]$ with $i\neq j$.
        \item \textbf{Views}: Let $\rho$ be the input to the $n$ parties. The view of parties in $T \subseteq [n]$ in round $k$, $\mathsf{View}_{T, k}(\rho)$, is the reduced density on all registers for parties in $T$ at the end of the $k$th round.
    \end{itemize}
\end{definition}

\begin{definition}[Secure Multiparty Quantum Computation (MPQC)~\cite{CGS02, Smith01}]
    \label[definition]{def:mpqc}
    $\Pi$ is an $n$-party, $t$-robust, and $t$-private MPQC protocol which computes a quantum circuit $\cC$ (with $n$ inputs and $n$ outputs) if we can define a ``real'' and ``ideal'' model with the following properties.

    \paragraph{Syntax.} The MPQC protocol can be written as in \cref{def:mpqc-syntax}.

    \paragraph{Real Model.} Given an adversary $\cA$ which can corrupt at most $t$ parties (but not abort), and an input configuration $\rho$:
    \begin{itemize}
        \item Every pair of parties are connected by perfect quantum and classical channels. 
        There is a classical authenticated broadcast channel.
        \item When $t < n/3$, the parties can perform classical multi-party computations.
        \item Initial configuration is joint state $\rho$ of $n+2$ systems: input system $\cI_i$ containing the parties' inputs for $i \in [n]$, $\cA$'s auxiliary input system $\cI_{aux}$, and an outside reference system $\cI_{ref}$; input can be an arbitrary quantum state, possibly entangling all these systems.
        \item Protocol interaction: all parties receive their input system $\cI_i$ and $\cA$ receives the state $\cI_{aux}$; $\cA$ chooses subset $C$ of size at most $t$ of parties to corrupt; $\cA$ has access to the state of the players in $C$ and controls what they send over the channels; $\cA$ may cause the cheaters' systems to interact arbitrarily; $\cA$ has no access to the state of the honest players and cannot intercept their communication; the reference system $\cI_{ref}$ is untouched during this process.
        \item Output: all parties produce output (honest parties output according to the protocol); system output by party $i$ is denoted $\cO_i$; adversary outputs an additional system $\cO_{aux}$.
        \item Output configuration: joint state of $\cO_1, \ldots, \cO_n$; $\cA$'s state $\cO_{aux}$; and the reference system $\cI_{ref}$. This state depends on $\cA$ and the initial configuration $\rho$, and is denoted $Real(\cA, \rho)$. (This configuration does not include any ancillary states or workspace used by honest parties, only the output specified by the protocol, all other parts of the honest parties' systems are ''traced out''.)
    \end{itemize}

    \paragraph{Ideal Model.} 
    Given an adversary $\cA$ which can corrupt at most $t$ parties (but not abort), and an input configuration $\rho$:
    \begin{itemize}
        \item There is a trusted (uncorruptable) third party $\mathcal{TTP}$ which receives no input.
        \item The communications model is the same as in the real model, except that every party is connected to $\mathcal{TTP}$ via a perfect (i.e. authentic, secret) quantum channel.
        \item Initial configuration: $n$ systems $\cI_i$ containing the parties' inputs, system $\cI_{aux}$, and system $\cI_{ref}$.
        \item The protocol proceeds as in the real model, except that the players may interact with $\mathcal{TTP}$.
        \item Output: same as in real model (final state of $\mathcal{TTP}$ is not included); the output configuration for adversary $\cA$ and initial configuration $\rho$ is denoted $Ideal(\cA, \rho)$.
    \end{itemize}
    Given a quantum circuit $\cC = U_T \ldots U_1$ with $n$ inputs and $n$ outputs:
    \begin{itemize}
        \item \underline{Input:} Each party gets an input system $I_i$.
        \item \underline{Input Sharing:} For each party $i$, party $i$ sends $I_i$ to $TTP$. If $TTP$ does not receive anything, then they broadcast ``Party $i$ is cheating'' to all parties. Otherwise, $TTP$ broadcasts ``Party $i$ is OK.''
        \item \underline{Computation:} $TTP$ evaluates the circuit $\cC$ on the inputs $I_i$. For all $i$ who cheated, $TTP$ creates $I_i$ in a known state (say $\ket{0}$).
        \item \underline{Output:} $TTP$ sends $i^{th}$ output to party $i$. Party $i$ outputs the system they receive from $TTP$.
    \end{itemize}

    \paragraph{Properties.}
    \begin{itemize}

        \item \textbf{Statistical Security:}
        There exists a quantum polynomial-time simulator $\Sim$ such that for all quantum unbounded-size circuits $\cA$ that corrupt at most $t$ parties and all sequences of input configurations $\{\rho_\lambda\}$ (possibly mixed or entangled), we have that for large enough $\lambda \in \bbN$: 
\begin{equation*}
            \| Real(1^\lambda, \cA, \rho_\lambda) - Ideal(1^\lambda, \Sim^{\cA}, \rho_\lambda) \|_1 \le \frac{1}{2^{\lambda/2 - 1}}.
        \end{equation*}

        \item \textbf{Statistical Correctness:} For all initial configurations $\{\rho_\lambda\}_{\lambda \in \bbN}$, there exists a negligible function $\negl(\cdot)$ such that for all security parameters $\lambda \in \bbN$ it holds that
        \begin{equation*}
            \|\mathsf{Output}(1^\lambda, \rho_\lambda) - \cC(\rho_\lambda)) \|_1 \le \negl(\lambda)
        \end{equation*}

        \item \textbf{Perfect $t$-Privacy:} Let $1 \le t \le n$ and $K$ be the number of rounds of the MPQC protocol. We say that $\Pi$ realizes $\cC$ with perfect $t$-privacy if there is a polynomial-size quantum circuit $\Sim$ such that 
for all input configurations $\{\rho_\lambda\}_{\lambda \in \bbN}$, all security parameters $\lambda \in \bbN$, all set of players $T \subseteq [n]$ where $|T| \le t$, and all rounds $k \in [K]$, it holds that
        \begin{equation*}
\mathsf{View}_{T,k}(1^\lambda, \rho_\lambda) \equiv \Sim(1^\lambda, T, \rho_\lambda\vert_T, \cC_T(\rho_\lambda), k)
        \end{equation*}
        where $\mathsf{View}_{T,k}(1^\lambda, \rho_\lambda)$ is defined in \cref{def:mpqc-syntax}.
    \end{itemize}
\end{definition}

We highlight that there are two notions concerning the privacy of the parties' inputs in the definition above, statistical security and perfect privacy. The key difference between these two notions lies in the adversarial model: in statistical security, simulation needs to be achieved against a malicious adversary, arbitrarily choosing the messages output in the protocol, while, in the privacy notion, the confidentiality of the inputs only needs to be protected against an adversary that generates all of its messages accordingly to the protocol description.
    
In the theorem below, we show that there exists a statistically secure protocol that also achieves perfect privacy; trading weaker security for stronger indistinguishability while increasing the threshold by a factor of $2$. Looking ahead, in the proof of our main protocol, we crucially rely on the perfect privacy of the MPQC.

\begin{theorem}[MPQC~\cite{CGS02, Smith01}]
    Let $n \in \bbN$ and $t < n/6$. There exists an $n$-party, statistically secure against $t$ corrupted parties, and perfectly $2t$-private MPQC protocol (as in \cref{def:mpqc}).
\end{theorem}
\begin{proof}[Proof (Sketch)] Since the work does not contain an explicit statement about $t$-privacy, we provide a sketch of the proof here. Before delving into the argument, it is useful to recall a high-level description of the main steps of the protocol. The protocol proceeds as follows:
    \begin{itemize}
        \item All parties share their input via a verifiable quantum secret sharing scheme.
        \item All computations are transversally over the shares, except for the measurement used in the degree reduction step. The result of this measurement is broadcast to all parties.
        \item To reconstruct the output, each party sends the share of their $i$-th output wire to the $i$-th party.
    \end{itemize}
    It is shown in~\cite[Lemma 2.23]{Smith01} that the measurement done in the degree-reduction step is independent of the inputs, and therefore broadcasting this value reveals no information about the inputs. The above described broadcasted value is the value that will be generated by the simulator $\Sim(1^\lambda, \dots, k)$ for the corresponding round $k$. Thus, what is left to be shown is that the secret sharing protocol is perfectly simulatable for $2t$ shares. Crucially, as mentioned above, the notion of $2t$-privacy only requires simulatability in the presence of honest parties (i.e., the analysis does not need to take into account corrupted parties), which is the reason why one can prove perfect, as opposed to statistical, privacy. It is shown in~\cite[Section 2.2.6]{Smith01} (considering only the case where the dealer D is honest) that the secret sharing is perfectly private, as long as at most $n - (n-2t) = 2t$ shares are revealed to the distinguisher. Informally, this is shown by arguing that the following procedures are equivalent:
\begin{itemize}
    \item Share the input state using the honest algorithm.
    \item Share the $\ket{0}$ state, then run the ideal interpolation on an arbitrary set of $n - 2t$ honest parties. Swap the input state sent by the trusted third party, and run the ideal interpolation circuit backwards.
\end{itemize} 
This equivalence follows by arguing that the verification of the verifiable secret sharing acts as the identity if the dealer D is honest.   
\end{proof}

\subsection{MPQC to Circuit-to-Hamiltonian}
\label{sec:mpqc-to-cth}

In the \cref{sec:qmaprotocol}, we will apply the Circuit-to-Hamiltonian reduction to the circuit simulating the MPQC interaction. Given an MPQC protocol $\Pi_{\mathcal{C}}$, we derive the corresponding quantum circuit as follows: The inputs for the parties are given as inputs to the circuits, along with sufficiently many ancillas $\ket{0}$. The gates of the circuit correspond to the unitaries $U_{i,k}$ (computing the ``next-message'' function for each party) acting on each party's workspace, along with SWAP gates that correspond to sending the message to another party.

To argue the indistinguishability of the resulting history state, if different MPQC inputs are used that lead to the same circuit evaluation, i.e., $\mathcal{C}(x^0_\lambda)=\mathcal{C}(x^1_\lambda)$, we introduce and prove the following helping lemma.

\begin{lemma}
    \label[lemma]{lem:mpqc-to-cth}    
Let $\mathcal{C}$ be a circuit computed by a perfectly $t$-private MPQC protocol $\Pi_{\mathcal{C}}$. Then, for all $\lambda\in \mathbb{N}$, the following holds. Let $x^0_\lambda$ and $x^1_\lambda$ be two (pure state) inputs such that $\mathcal{C}(x^0_\lambda)=\mathcal{C}(x^1_\lambda)$ and $x_\lambda^0\vert_T = x_\lambda^1\vert_T =x_\lambda\vert_T$, where $T$ is a set of corrupted parties with $|T|\leq t$. Define $\phi^0_{hist,T}$ and $\phi^1_{hist,T}$ to be the reduced densities on the registers corresponding to the parties in $T$ and the clock register, of the history states generated by executing $\Pi_{\mathcal{C}}(x^0_{\lambda})$ and $\Pi_{\mathcal{C}}(x^1_\lambda)$, respectively.
Then 
    \[
    \phi^0_{hist,T} \equiv \phi^1_{hist,T}.
    \]
\end{lemma}
\begin{proof}
    First observe that, by the $t$-privacy of the MPQC protocol $\Pi_{\mathcal{C}}$, we have that
    \[
    \mathsf{View}_{T,k}(1^\lambda, x_\lambda^0) \equiv \Sim(1^\lambda, T, x_\lambda\vert_T, \cC_T(x_\lambda), k) \equiv \mathsf{View}_{T,k}(1^\lambda, x_\lambda^1)
    \]
    for any step $k$ of the computation, where the view $\mathsf{View}_{T,k}(1^\lambda, x^b_\lambda)$ is defined as in~\cref{def:mpqc-syntax}. Recall that the Circuit-to Hamiltonian reduction results in a state that contains a superposition of all steps of the computation. Note that
    \[
    \ket{\phi_{hist}^b}=\frac{1}{\sqrt{K+1}}\sum_{\kappa=1}^K U'_\kappa\dots U'_1\ket{x_\lambda^b,0}\otimes\ket{\kappa} 
    =\frac{1}{\sqrt{K+1}}\sum_{\kappa=1}^K \mathsf{View}_{\kappa}(1^\lambda, x^b_\lambda)\otimes\ket{\kappa} 
    \]
    where $U_i'$ denotes the unitary applied at the $i$-th step of the circuit. Let us consider each state $\mathsf{View}_{\kappa}(1^\lambda, x^b_\lambda)\otimes\ket{\kappa}$, for all $\kappa$. When tracing out all registers not in $T$, the argument above establishes that the two reduced densities (for $b=0$ and $b=1$) are identical. Thus, by Uhlmann theorem, there exists a unitary $V_{\kappa}$ acting on the traced out system such that
    \[
    (V_{\kappa} \otimes \mathsf{Id}_T) \mathsf{View}_{\kappa}(1^\lambda, x^0_\lambda)\otimes\ket{\kappa}  = \mathsf{View}_{\kappa}(1^\lambda, x^1_\lambda)\otimes\ket{\kappa} .
    \]
    We can assume without loss of generality that all honest parties of the MPQC keep track of the step number in one of its internal registers and we refer to one such register (that is not part of $T$) as $R$. Furthermore, we can assume that $V_\kappa = V_\kappa'\otimes \mathsf{Id}_R$, since the register $R$ is in the state $\ket{\kappa}$ for both $\mathsf{View}_{\kappa}(1^\lambda, x^0_\lambda)$ and $\mathsf{View}_{\kappa}(1^\lambda, x^1_\lambda)$. This allows us to define $V = \sum_{\kappa} V_\kappa' \otimes \ketbra{\kappa}_R$, as the controlled application of $V_\kappa$ on $R$. Then we have that 
    \[
    (V \otimes \mathsf{Id}_T) \ket{\phi_{hist}^0} = \ket{\phi_{hist}^1}
    \]
    Observe that $V$ does not act on any of the $T$ registers, which implies that $\phi^0_{hist,T} \equiv \phi^1_{hist,T}$.
\end{proof}

\section{Superposition-Attack Secure, Zero-Knowledge Argument for $\NP$}
\label{sec:zk-np}

\subsection{Secret Sharing Scheme Secure Against Superposition Attacks~\cite{DamgardFNS13}}

\begin{definition}[Party views]
Let $P_1, \ldots, P_n$ be parties in a secret sharing scheme $S$ where $s \in \bbS$ is the secret shared using randomness $r \in \cR$. Each party, $P_i$, receives a share $v_i(s; r) \in \zo^k$, also called its private view. For $A \subset [n]$, let $v_A(s; r) = \{v_i(s; r)\}_{i \in A}$ be the string containing the concatenation of views for
parties $P_i$ with $i \in A$.
\end{definition}

\begin{definition}[Perfect security]
\label[definition]{def:g-attacks}
Let $G$ be a family of subsets of $2^{[n]}$. A secret sharing scheme $S$ is perfectly secure
against classical $G$-attacks if for any $A \in G$, the distribution of shares $v_A(s; r) \in \zo^t$ does not depend on $s$.
\end{definition}

\begin{definition}[Adversary structure]
\label[definition]{def:adv-str}
The adversary structure of a secret sharing scheme is the maximal $G \subset 2^{[n]}$ for which the scheme is
perfectly secure against classical $G$-attacks (\cref{def:g-attacks}).
\end{definition}

\begin{definition}[Perfect security against superposition attacks]
\label[definition]{def:ps-sup}
A secret sharing scheme $S$ is perfectly secure against superposition $F$-attacks if, and only if, for all unitary matrices, \begin{equation*}
    U^{adv, F}_{query} \: : \: \cH_{env} \otimes \cH_{query} \to \cH_{env} \otimes \cH_{query}
\end{equation*}
and all possible pairs of inputs, $s, s' \in \bbS$ it holds that $\rho^{adv,F}_s = \rho^{adv,F}_{s'}$, where for $s^* \in \{s, s'\}$
\begin{align*}
    \rho^{adv,F}_{s^*} &= \sum_{r\in\cR}p_r\ket{\psi_{adv,F,s^*}}\allowbreak\bra{\psi_{adv,F,s^*}}, \text{ and} \\
    \ket{\psi_{adv,F,s^*}}& =
    U^{adv, F}_{query} \left( \sum_x \alpha_x \ket{x}_e \otimes \ket{0,0}_q\right)\\
    &=\sum_{x,A\in F,a\in\bits^t}\alpha_{x,A,a}\ket{x}_e\otimes\ket{A,a\oplus v_A(s^*,r)}_q.
\end{align*}
Here, $\cH_{env}$ is the environment in which the adversary can store auxiliary information. The dimension of this register is arbitrary, but finite. In the $\cH_{query}$ register, the adversary submits its queries and receives the corresponding responses from the oracle.
\end{definition}

\begin{remark}
    \label{rem:sss}
    If $S = (\mathsf{Share}, \mathsf{Reconstruct})$ is $t$-secure (as defined in \cref{def:sss}), then 
$S$ is secure against classical $G$-attacks where $G$ is the family of all subsets of size $t$ (\cref{def:g-attacks}).
\end{remark}

\begin{theorem}[{\cite[Theorem 1]{DamgardFNS13}}]
\label{thm:dfns}
Let $S$ be a secret sharing scheme with adversary structure $G$. If $F^2 \subseteq G$ where $F^2 \triangleq \{A \: \vert \: \exists B, C \in F \: :\: A = B \cup C\}$, then $S$ is perfectly secure against superposition $F$-attacks (\cref{def:ps-sup}).
\end{theorem}
For completeness, we refer the reader to \cref{app:dfns-thm1} for a full proof.

\subsection{Our Protocol}

We describe our protocol in \cref{fig:zk-for-np}. Our protocol draws upon the prior works of~\cite{IshaiKOS09} and~\cite{DamgardFNS13}.

\begin{figure}[!ht]
\begin{framed}
\centering
\begin{minipage}{1.0\textwidth}
\begin{center}
    \underline{Superposition-Attack Secure Zero-Knowledge Proof for $\NP$ from MPC}
\end{center}

\vspace{2mm}

Let $(\Gen_H, \Gen_B, \DualCom)$ be a dual-mode commitment scheme.
Let $(\Gen, \Com)$ be a perfectly binding, computationally hiding commitment scheme.
Let $\Pi$ be an $n$-party, perfectly correct, perfectly $t$-private, perfectly $t$-robust MPC protocol. Let $\cR$ be an $\NP$ relation.

\vspace{0.2cm}

\noindent {$\underline{\textsc{Setup}(1^\lambda)}$}:
Let $\crs = (\ck_B, \ck, c)$ where two commitment keys and a commitment are sampled as follows: $\ck_B \gets \Gen_B(1^\lambda)$, $\ck \gets \Gen(1^\lambda)$, and $c \gets \Com(\ck, 0)$. Outputs $\crs$.

\vspace{0.2cm}

Let $\cR' = \left\{ ((x, c, \ck), (w, r)) \sthat (x, w) \in \cR \: \vee \: c \equiv \Com(\ck, 1; r) \right\}$. Let $\cC_\cR'$ be a circuit that on input $(x', (w_1, r_1), \ldots, (w_n, r_n))$ computes $\cR'(x', \xor_{i \in [n]} (w_i, r_i))$.
Let $(x, w) \in \cR$.

\vspace{0.3cm}

\noindent {$\underline{\sP_1(\crs, x, w)}$}:
\begin{itemize}
    \item Samples $r \gets \{0,1\}^{\lambda}$ uniformly at random. Let $m$ be the size of witnesses to $\cR$. Sample witness shares $w_1', \ldots, w_n' \in \{0, 1\}^m \times \{0, 1\}^\lambda$ such that $\xor_{i \in [n]} w_i' = (w, r)$. 
    \item Emulates an execution of $\Pi$ for $\cC_{\cR'}$ on input $(x', w_1', \ldots, w_n')$. Defines $V_1, \ldots, V_n$ to be the views of the $n$ players.
        \item Computes commitments $\alpha_i = \DualCom(\ck_B, V_i; r_i)$ with uniformly random $r_i$ for $i \in [n]$. Defines openings $\zeta_i = (V_i, r_i)$ for $i \in [n]$. 
    \item Outputs $(\alpha = (\alpha_1, \ldots, \alpha_n), \zeta = (\zeta_1, \ldots, \zeta_n))$.
\end{itemize}

\noindent {$\underline{\sV_1(\crs, x, \alpha)}$}:
Outputs a set of $t/2$ parties: $\beta = (\beta_1, \ldots, \beta_{t/2})$ where $\beta_i \in [n]$ for $i \in [t/2]$.

\vspace{0.3cm}

\noindent {$\underline{\sP_2(\crs, x, w, \alpha, \beta, \zeta)}$}:
Outputs $\gamma = \{\zeta_{\beta_i}\}_{i \in [t/2]}$ with $\beta = (\beta_1, \ldots, \beta_{t/2})$ and $\zeta = (\zeta_1, \ldots, \zeta_n)$.

\vspace{0.3cm}

\noindent {$\underline{\sV_2(\crs, x, \alpha, \beta, \gamma)}$}:
Outputs $1$ if and only if the following checks pass:
\begin{itemize}
    \item The openings are valid, i.e.\ $\alpha_{\beta_i} \equiv \DualCom(\ck_B, V_{\beta_i}; r_{\beta_i})$ for $i \in [t/2]$.
    \item For $i \in [t/2]$, view $V_{\beta_i}$ has the $\beta_i$th party output $1$.
    \item The views $V_{\beta_i}$ and $V_{\beta_j}$ are consistent with respect to $\Pi$ and $x$ for $i, j \in [t/2]$.
\end{itemize}
Else, outputs $0$.

\vspace{0.1cm}

\end{minipage}
\end{framed}
\caption{Superposition-Secure MPC in the Head for $\NP$}
\label{fig:zk-for-np}
\end{figure}

\subsection{Analysis}

\begin{theorem}
    Let $n \in \bbN$ and $t < n/3$.
    Let a dual-mode commitment $(\Gen_H, \Gen_B, \DualCom)$ (\cref{def:dm-com}), a perfectly binding commitment $(\Gen, \Com)$ (\cref{def:com}), and a $n$-party, perfectly correct, perfectly $t$-private, perfectly $t$-robust MPC protocol $\Pi$ (\cref{def:mpc}) be given. 
Then, the protocol in \cref{fig:zk-for-np} is a statistically sound, computationally zero-knowledge against superposition-attacks proof for $\NP$ in the common reference string model.
\end{theorem}

\begin{proof}
    ~\paragraph{Completeness.}
    This follows from the perfect binding of the commitment scheme and the MPC protocol.

    \paragraph{Statistical Soundness.}
    Our proof closely follows the proof from~\cite[Theorem 4.1]{IshaiKOS09}.
    Let a language $\cL \in \NP$, a polynomial $p(\cdot)$, an unbounded-size quantum circuit $\cA$, and an instance $x \not\in \cL$ be given such that
    \begin{align}
        &\Pr_{\substack{\crs \gets \Setup(1^\lambda)}}[\langle \cA_{\lambda}, \sV \rangle(x, \crs) = 1]  \ge \frac{1}{p(\lambda)}. \label{eq:break-sound-np}
    \end{align}
    
$\cA$ is unable to break the perfect binding property
with respect to the $(\Gen, \Com)$ commitment.
As such, we know that there is no private input for each party which would cause an honest execution of $\cC_{\cR'}$ to output $1$.
    We also have that $\cA$ is unable to break the perfect binding property with respect to the dual-mode commitment.
    This leads us to having only two cases in this scenario: the prover corrupts either less than (or equal to) $t$ or greater than $t$ parties in the MPC protocol. We show contradictions in both of these two scenarios.
    \begin{itemize}
        \item Say we have that $\cA$ corrupts less than (or equal to) $t$ parties in the MPC protocol and manages to succeed in the event from \cref{eq:break-sound-np} with advantage $\frac{1}{2p(\lambda)}$. 
When the instance $x \not\in \cL$, then, by the $t$-robustness of the MPC protocol, no uncorrupted party will output $1$. As such, for the verifier to accept, it must necessarily be the case that the verifier chooses to open only corrupted parties. This happens with probability at most $(\frac{t}{n})^{t/2}$, which we calculate with replacement, which gives a negligible upperbound on the soundness error in this case.

        \item Say instead that $\cA$ corrupts greater than $t$ parties in the MPC protocol and manages to succeed in the event from \cref{eq:break-sound-np} with advantage $\frac{1}{2p(\lambda)}$. For the verifier to accept, the verifier cannot pick any pair of parties with ``inconsistent'' views. All pairs of inconsistent parties have at least one party in the set of at least $t$ corrupted parties.
        So, there must be strictly greater than $t/2$ independent~\footnote{Here by ``independent'' we mean that the two parties in the pair are inconsistent with each other irrespective of their relationship with other parties.} pairs of inconsistent parties. For simplicity of computing the soundness error, we convert the notion of picking inconsistent pairs to picking a set of parties without inconsistencies.
        
        We count the number of ways that the verifier could avoid picking even one of these pairs of parties.
        For our count, we will assume the worst case, where there are only $t/2$ independent pairs.
        We will enumerate over the number of parties, $k \in [t/2]$, that are picked from the set of those within an independent pair. There are $\binom{t/2}{k}$ such ways to choose a pair and $2$ options of party to pick for each choice, leading to a count of $\binom{t/2}{k} 2^k$. Now, we must choose the remaining $t/2 - k$ from the $n - t$ parties which do not have a pair, leading to a count of $\binom{n-t}{t/2-k}$. As such, we get a union bound on probability that the verifier does not pick even one of these pairs of parties,
        \begin{equation*}
            \left. \sum_{k \in [t/2]} \binom{n - t}{t/2 - k}\binom{t/2}{k} 2^k \right/ \binom{n}{t/2},
        \end{equation*}
        which yields a negligible upperbound on the soundness error in this setting.

    \end{itemize}
    Since we have negligible error in either scenario, we have an overall negligible soundness error.

    \paragraph{Superposition-secure Computational Zero-Knowledge.}
    To model superposition access of the adversary, we consider a classical algorithm $\sP'_2$ that has values $(\crs, x, w, \alpha, \zeta)$ hardcoded inside and takes as an input only $\beta$ and then, internally, runs $\sP_2(\crs,\allowbreak x, w, \alpha, \beta, \zeta)$. Given the classical algorithm for $\sP'_2$, there also exists a unitary $U_{\sP'_2}$ implementing $\sP'_2$ quantumly. When the verifier now submits a superposition query $\sum_\iota \ket{\beta_\iota}$, it is possible to compute the reply, an opening over all the challenges contained in this superposition, by applying the unitary $U_{\sP'_2}$ controlled on the challenge $\sum_\iota \ket{\beta_\iota}$. More precisely, we apply $U_{\sP'_2}\sum_\iota \ket{\beta_\iota}\ket{0}_\regM=\sum_\iota \ket{\beta_\iota}\ket{\gamma_\iota}_\regM$ and then return the register $\regM$ as a reply to the verifier.

    We define a two-part simulator $\Sim=(\Sim_1, \Sim_2)$. The first algorithm $\Sim_1$ takes as input an instance $x$:
    \begin{itemize}
        \item Computes $\crs = (\ck_H, \ck, c)$ by sampling two commitment keys $\ck_H \gets \Gen_H(1^\lambda)$, $\ck \gets \Gen(1^\lambda)$, and computing the commitment $c = \Com(\ck, 1; r)$ with uniform randomness $r$.
\item Samples $w\gets\bits^m$ uniformly at random. Samples witness shares $w_1', \ldots, w_n' \in \{0, 1\}^m \times \{0, 1\}^\lambda$ such that $\xor_{i \in [n]} w_i' = (w, r)$. Sets $x' = (x, c, \ck)$.
\item Emulates an execution of $\Pi$ for $\cC_{\cR'}$ on input $(x', w_1', \ldots, w_n')$. Defines $V_1, \ldots, V_n$ to be the views of the $n$ players.
        \item Computes commitments $\alpha_i = \DualCom(\ck_H, V_i; r_i)$ with uniformly random $r_i$ for $i \in [n]$. Defines openings $\zeta_i = (V_i, r_i)$ for $i \in [n]$.
\item Outputs $(\crs, \alpha = (\alpha_1, \ldots, \alpha_n), \zeta = (\zeta_1, \ldots, \zeta_n))$.
    \end{itemize}
    The second algorithm $\Sim_2$ takes $(\crs, x, w, \alpha, \sum_\iota \ket{\beta_\iota}, \zeta)$ as input, where $\sum_\iota \ket{\beta_\iota}$ may be a superposition query, hardcodes $(\crs, x, w, \alpha, \zeta)$ into $U_{\sP_2'}$, and outputs $U_{\sP_2'}(\sum_\iota \ket{\beta_\iota})$.

    Some of the hybrids for the proof of our construction are inefficient. For example the third hybrid in which the commitment schemes are opened an inefficient way. We highlight that this inefficiency is only required to prove the indistinguishability between some of the hybrids and that the final simulator described above is efficient.

    Let a relation $\cR$ and a quantum polynomial-size quantum circuit $\cA$ that can send superposition queries be given.
    We define the following series of hybrids, for any $\lambda \in \bbN$ and instance-witness pair $(x, w) \in \cR$ where $|x| = \lambda$, which we will argue are all indistinguishable:
    \begin{itemize}
        \item $\cH_0$: Same as the real world. 
        
        $\crs \gets \Setup(1^\lambda)$, $(\alpha, \zeta) \gets \sP_1(\crs, x, w)$, $(\sum_\iota \ket{\beta_\iota}, \xi) \gets \cA(\crs, x, \alpha)$, hardcode $U_{\sP_2'}$ with $(\crs, x, w, \alpha, \zeta)$, compute $U_{\sP'_2}(\sum_\iota \ket{\beta_\iota}\ket{0}_\regM)=\sum_\iota \ket{\beta_\iota}\ket{\gamma_\iota}_\regM$, and where register $\regM$ is output.

        \item $\cH_1$: 
        Same as $\cH_0$ except the commitment in the $\crs$ is a commitment to $1$. 
        
        $\ck_B \gets \Gen_B(1^\lambda)$, $\ck \gets \Gen(1^\lambda)$, $c \gets \Com(\ck, 1)$, $\crs = (\ck_B, \ck, c)$, $(\alpha, \zeta) \gets \sP_1(\crs, x, w)$, $(\beta, \xi) \gets \cA(\crs, x, \alpha)$, hardcode $U_{\sP_2'}$ with $(\crs, x, w, \alpha, \zeta)$, compute $U_{\sP'_2}(\sum_\iota \ket{\beta_\iota}\ket{0}_\regM)=\sum_\iota \ket{\beta_\iota}\ket{\gamma_\iota}_\regM$, and where register $\regM$ is output.
        
        The indistinguishability between this and the previous hybrid follows from the hiding of the commitment scheme $\Com$. 
        We prove the indistinguishability of $\cH_0$ and $\cH_1$ in \cref{claim:h0-to-h1}.

        \item $\cH_2$: 
        Same as $\cH_1$ except the dual mode commitment is in hiding mode. 
        
        $\ck_H \gets \Gen_H(1^\lambda)$, $\ck \gets \Gen(1^\lambda)$, $c \gets \Com(\ck, 1)$, $\crs = (\ck_H, \ck, c)$, $(\alpha, \zeta) \gets \sP_1(\crs, x, w)$, $(\sum_\iota \ket{\beta_\iota}, \xi) \gets \cA(\crs, x, \alpha)$, hardcode $U_{\sP_2'}$ with $(\crs, x, w, \alpha, \zeta)$, compute $U_{\sP'_2}(\sum_\iota \ket{\beta_\iota}\ket{0}_\regM)=\sum_\iota \ket{\beta_\iota}\ket{\gamma_\iota}_\regM$, and where register $\regM$ is output.
        
        The indistinguishability between this and the previous hybrid follows from the indistinguishability of modes from the dual-mode commitment scheme.
        We prove the indistinguishability of $\cH_1$ and $\cH_2$ in \cref{claim:h1-to-h2}.
         
        \item $\cH_3$: It works the same as $\cH_2$ except the prover's first message is replaced with commitments to $0$, and the binding of the commitments are broken to open them to the prover's third message. This hybrid runs in exponential time.

        For this hybrid, and the following, we define the following polynomial-size classical circuit $\cB$ which is hardcoded with $\{(V_i, r_i)\}_{i \in [n]}$: output $\{(V_{\beta_i}, r_{\beta_i})\}_{i \in [t/2]}$.
        This classical circuit can be implemented as a unitary $U_{\cB}$.
        
        $\ck_H \gets \Gen_H(1^\lambda)$, $\ck \gets \Gen(1^\lambda)$, $c \gets \Com(\ck, 1)$, $\crs = (\ck_H, \ck, c)$, $(\cdot, \{(V_i, \cdot)\}_{i \in [n]}) \gets \sP_1(\crs, x, w)$, $\alpha = (\alpha_1, \ldots, \alpha_n)$ where $\alpha_i = \DualCom(\ck_H, 0)$, $(\sum_\iota \ket{\beta_\iota}, \xi) \gets \cA(\crs, x, \alpha)$, computes valid openings $(r_1, \ldots, r_n)$ to $(V_1, \ldots, V_n)$ (by running in unbounded-time), hardcodes $U_{\cB}$ with $\{(V_i, r_i)\}_{i \in [n]}$, and computes $U_{\cB}(\sum_\iota \ket{\beta_\iota}\ket{0}_\regM)=\sum_\iota \ket{\beta_\iota}\ket{\gamma_\iota}_\regM$ where $\regM$ is output.
        
        The indistinguishability between this and the previous hybrid follows from the statistical hiding property of the dual-mode commitment scheme. 
        We prove the indistinguishability of $\cH_2$ and $\cH_3$ in \cref{claim:h2-to-h3}.

        \item $\cH_4$: Same as $\cH_3$ except the binding of the commitments are broken to open them to the simulator's third message. This hybrid also runs in exponential time.

        See the definition of $U_{\cB}$ in the previous hybrid.
        
$(\crs, \cdot, \{(V_i, \cdot)\}_{i \in [n]}) \gets \Sim_1(x)$, $\alpha = (\alpha_1, \ldots, \alpha_n)$ where $\alpha_i = \DualCom(\ck_H, 0)$, $(\sum_\iota \ket{\beta_\iota}, \xi) \gets \cA(\crs, x, \alpha)$, computes valid openings $(r_1, \ldots, r_n)$ to $(V_1, \ldots, V_n)$ (by running in unbounded-time), hardcodes $U_{\cB}$ with $\{(V_i, r_i)\}_{i \in [n]}$, compute $U_{\cB}(\sum_\iota \ket{\beta_\iota}\ket{0}_\regM)=\sum_\iota \ket{\beta_\iota}\ket{\gamma_\iota}_\regM$ where register $\regM$ is output.

        The indistinguishability between this and the previous hybrid follows from the $t$-privacy of the underlying MPC protocol.  We prove the indistinguishability of $\cH_3$ and $\cH_4$ in \cref{claim:h3-to-h4} using the results of~\cref{lem:t-out-of-n-classical}.
        
        \item $\cH_5$: This hybrid corresponds to the ideal world. It is the same as $\cH_4$ except that the dual-mode commitments are generated using the views of the different parties.
        
        $(\crs, \alpha, \zeta) \gets \Sim_1(x)$, $(\sum_\iota \ket{\beta_\iota}, \xi) \gets \cA(\crs, x, \alpha)$, and $\Sim_2(\crs, x, w, \alpha, \sum_\iota \ket{\beta_\iota}, \zeta)$ outputs register $\regM$. 
        
        The indistinguishability between this and the previous hybrid follows by, again, relying on the statistical hiding property of the dual-mode commitment scheme. The proof of this transition proceeds similarly to the proof of \cref{claim:h2-to-h3}.
    \end{itemize}

To prove our next claim, we construct a reduction that breaks the following property which is implied by the computational hiding property in \cref{def:com}: instead of quantifying over all messages $m$ and $m'$, the distinguisher gets to choose $m$ and $m'$ before it receives the commitment key.
    To see why this implication holds, imagine a distinguisher $\cD$ broke the property we wrote above. We can define a reduction that receives $m$ and $m'$ from $\cD$ as the first message in its interaction. The reduction then sets the parameters to the game that it will play with the challenger from \cref{def:com} to be with respect to $m$ and $m'$. Now the reduction's success probability will be the same as the success probability of $\cD$. This shows the implication.

    \begin{claim}
        \label[claim]{claim:h0-to-h1}
        Let $(\Gen,\Com)$ be a computationally hiding commitment scheme, then the hybrids $\cH_0$ and $\cH_1$ are computationally indistinguishable.
    \end{claim}
    \begin{proof}
        We have that $\cH_0$ and $\cH_1$ are indistinguishable by the computational hiding of the plain commitment scheme. Say, for contradiction that there existed a quantum polynomial-size quantum circuit $\cD$ that given input $(\crs, x, \alpha, \sum_\iota \ket{\beta_\iota, \gamma_\iota}, \zeta)$ could distinguish between $\cH_0$ and $\cH_1$. 

        We define an efficient reduction, hardwired with the instance-witness pair $(x, w)$, that receives $\ck \gets \Gen(1^\lambda)$ from the challenger. The reduction sends $0$ and $1$ to the challenger. The challenger sends back $c^*$. The reduction then samples the remainder of the common reference string $\crs = (\ck_B, \ck, c^*)$ as follows: samples $\ck_B \gets \Gen_B(1^\lambda)$. The reduction computes $(\alpha, \zeta) \gets \sP_1(\crs, x, w)$.
        The reduction then sends $(\crs, \alpha)$ to $\cA$.
        The reduction receives $\sum_\iota \ket{\beta_\iota}$ from $\cA$ and then, to compute the reply, the reduction hardcodes $U_{\sP_2'}$ with $(\crs, x, w, \alpha, \zeta)$ and computes $U_{\sP'_2}(\sum_\iota \ket{\beta_\iota}\ket{0})=\sum_\iota \ket{\beta_\iota}\ket{\gamma_\iota}$.
        Lastly, the reduction outputs the result of $\cD(\crs, x, \alpha, \sum_\iota \ket{\beta_\iota, \gamma_\iota}, \zeta)$.

        Our reduction will preserve the distinguishing advantage of $\cD$. However, this contradicts the property we defined above this claim. Since this property is implied by the computational hiding property in \cref{def:com}, we contradict the computational hiding of the commitment. As such, we must have that $\cH_0$ and $\cH_1$ are indistinguishable except with negligible probability.
    \end{proof}

    \begin{claim}
        \label[claim]{claim:h1-to-h2}
        Let $(\Gen_H,\Gen_B,\DualCom)$ be a dual-mode commitment scheme, then the hybrids $\cH_1$ and $\cH_2$ are computationally indistinguishable.
    \end{claim}
    \begin{proof}
        $\cH_1$ and $\cH_2$ are indistinguishable by the computational indistinguishability of hiding and binding modes for the dual-mode commitment. Say, for contradiction that there existed a quantum polynomial-size quantum circuit $\cD$ that given input $(\crs, x, \alpha, \sum_\iota \ket{\beta_\iota, \gamma_\iota}, \xi)$ could distinguish between $\cH_1$ and $\cH_2$. 

        We define an efficient reduction, hardwired with the instance-witness pair $(x, w)$, that receives $\ck^*$ from the challenger either sampled according to $\Gen_H$ or $\Gen_B$. The reduction then samples the remainder of the common reference string $\crs = (\ck^*, \ck, c)$ as follows: samples $\ck \gets \Gen(1^\lambda)$ and $c \gets \Com(\ck, 1)$. The reduction computes $(\alpha, \zeta) \gets \sP_1(\crs, x, w)$.
        The reduction then sends $(\crs, \alpha)$ to $\cA$.
        The reduction receives $(\sum_\iota \ket{\beta_\iota}, \xi)$ from $\cA$ and then, to compute the reply, the reduction hardcodes $U_{\sP_2'}$ with $(\crs, x, w, \alpha, \zeta)$ and computes $U_{\sP'_2}(\sum_\iota \ket{\beta_\iota}\ket{0})=\sum_\iota \ket{\beta_\iota}\ket{\gamma_\iota}$. 
Lastly, the reduction outputs the result of $\cD(\crs, x, \alpha, \sum_\iota \ket{\beta_\iota, \gamma_\iota}, \xi)$.

        Our reduction will preserve the distinguishing advantage of $\cD$. However, this contradicts the computational indistinguishability of keys for the dual-mode commitment scheme. As such, we must have that $\cH_1$ and $\cH_2$ are indistinguishable except with negligible probability.
    \end{proof}

    \begin{claim}
        \label[claim]{claim:h2-to-h3}
        Let $(\Gen_H,\Gen_B,\DualCom)$ be a dual-mode commitment scheme, then the hybrids $\cH_2$ and $\cH_3$ are statistically indistinguishable.
    \end{claim}
    \begin{proof}
        $\cH_2$ and $\cH_3$ are indistinguishable by the statistical hiding property of the dual-mode commitment. We argue this by switching each commitment one at a time and using a union bound to get an overall negligible distinguishing probability. Say we switch the first commitment, then the second, and so on until the $n$'th commitment. Let $i \in [n]$ be chosen arbitrarily. Say, for contradiction that there existed an unbounded-time distinguisher $\cD$ that given input $(\crs, x, \alpha, \beta, \gamma, \xi)$ could distinguish whether the $i$'th commitment was a commitment to a view or $0$.

        We define a computationally unbounded reduction, hardwired with the instance-witness pair $(x, w)$ and index $i$, that receives $\ck_B \gets \Gen_B(1^\lambda)$ from the challenger. The reduction then samples the remainder of the common reference string $\crs = (\ck_B, \ck, c)$ as follows: samples $\ck \gets \Gen(1^\lambda)$ and $c \gets \Com(\ck, 1)$. The reduction computes $(\alpha, \zeta) \gets \sP_1(\crs, x, w)$.
        The reduction then sends two messages $V_i$ (taken from $\alpha_i$) and $0$ to the challenger.
        The challenger sends back $c^*$.
        The reduction then sends $(\crs, \alpha')$ to $\cA$ where $\alpha' = (\DualCom(\ck_B, 0), \ldots, \DualCom(\ck_B, 0), c^*, \alpha_{i+1}, \ldots, \alpha_n)$.
        The reduction receives $(\sum_\iota \ket{\beta_\iota}, \xi)$ from $\cA$. 
In the next step, the reduction runs in unbounded time to find openings $(r'_1,\dots,r'_n)$ such that all the commitments $(\alpha_1,\dots,\alpha_n)$ open to $(V_1,\dots,V_n)$. 
Afterwards, to compute the reply, the reduction hardcodes $U_\cB$ with $\{(V_i, r_i')\}_{i \in [n]}$ and computes $U_{\cB}(\sum_\iota \ket{\beta_\iota}\ket{0})=\sum_\iota \ket{\beta_\iota}\ket{\gamma_\iota}$.
Lastly, the reduction outputs the result of $\cD(\crs, x, \alpha, \sum_\iota \ket{\beta_\iota, \gamma_\iota}, \xi)$.

        Our reduction will preserve the distinguishing advantage of $\cD$. However, this contradicts the statistical hiding of the commitment. As such, we must have that $\cH_2$ and $\cH_3$ are indistinguishable except with negligible probability.
        We have a similar argument for the indistinguishability of $\cH_4$ and $\cH_5$. The main difference is that we sample the common reference string and views of the parties according to $\Sim_1$.
    \end{proof}

    Before proceeding with the proof of~\cref{claim:h3-to-h4}, we cast the first two steps of the prove procedure $\sP_1(\crs,x,w)$, together with some parts of the CRS, as the sharing procedure of a secret sharing scheme for a secret $w\in\bits^m$. More precisely:
    \begin{description}
        \item[$\Setup(1^\secpar)$:] Generate a commitment key $\ck \gets \Gen(1^\lambda)$ for a perfectly binding, computationally hiding commitment scheme and a commitment $c \gets \Com(\ck, 1)$. Output $\crs = (\ck, c)$.
    
        \item[$\share(w)$:] Sample $r\gets\bits^\lambda$ as well as $w_1,\dots,w_{n-1}\gets\bits^m$ and $r_1,\dots,r_{n-1}\gets\bits^\lambda$ uniformly at random and set $(w_n,r_n)\coloneqq (w_1,r_1)\oplus\dots\oplus (w_{n-1},r_{n-1})\oplus(w,r)$. Emulate an execution $\Pi$ for $\cC_{\cR'}$ on input $(x', w_1', \ldots, w_n')$ where $w'_i\coloneqq(w_i,r_i)$ and let $s_1, \ldots, s_n$ to be the views of the players $P_1,\dots,P_n$. Output $\{s_i\}_{i\in[n]}$ as the set of shares.

        \item [$\rec(\crs,\{s_i\}_{i\in[n]})$:] Parse the shares $\{s_i\}_{i\in[n]}$ as the views of the parties $P_i$ for all $i\in[n]$. Then, potentially using $\crs$, reconstruct $w'_i$ from the view of $P_i$ for all $i\in[n]$. Compute $(w,r)=\oplus_{i\in[n]}w_i'$ and output $w$ as the secret.
    \end{description}

    In the next step, we prove that the presented secret sharing scheme achieves perfect $t$-out-of-$n$ security.

    \begin{claim}
        \label[claim]{lem:t-out-of-n-classical}
        Let $\Pi$ be a perfectly $t$-private MPC protocol, then the above secret sharing scheme is perfectly secure against classical $G$ attacks where $G$ contains all subsets of $[n]$ of size $t$.
    \end{claim}
    \begin{proof}
        To prove this lemma, it suffices to prove that the above described secret sharing scheme is a $t$-out-of-$n$ secret sharing scheme, following~\cref{rem:sss}. We prove this using the following hybrids:
        \begin{description}
            \item[Hybrid $\cH_0$:] This hybrid corresponds to the case where the secret sharing scheme is executed using the secret $w$.
            
            \item[Hybrid $\cH_1$:] This hybrid corresponds to the setting where the messages of the MPC protocol $\Pi$ are generated using the shares of the opening of the commitment $c$ as an input instead of honestly generated using the shares of the witness $w$. The perfect indistinguishability between this and the previous hybrid follows from the perfect $t$-privacy of the MPC protocol $\Pi$.

            \item[Hybrid $\cH_2$:] In this hybrid, the shares $w_1,\dots,w_n$ are sampled as a secret sharing of $0$ instead of the secret $w$. This hybrid is perfectly indistinguishable from the previous hybrid due to the security of the secret sharing scheme $w_1,\dots,w_n$. This hybrid corresponds to an execution of the sharing algorithm that does not rely on the witness $w$ and, therefore, concludes the proof.

\end{description}

        \paragraph{Indistinguishability of hybrids $\cH_0$ and $\cH_1$:} We prove the indistinguishability of the hybrids $\cH_0$ and $\cH_1$ using the perfect $t$-privacy of the MPC protocol. We do this by describing an adversary $\cB$ that interacts with the adversary $\cA$ of the secret sharing scheme. In the first step, the adversary $\cB$ generates the $\crs$ as described in the setup procedure and sends it to $\cA$, afterwards, it receives the secret $w$ as well as a set of indices $T$ from the adversary $\cA$. In the next step, the adversary $\cB$ samples $r_1,\dots,r_n$ such that $r\coloneqq\oplus_{i\in[n]}r_i$ where $r$ is the opening of the commitment of the $\crs$. Furthermore, the adversary samples $w_1,\dots,w_{n-1}\gets\bits^m$ uniformly at random and sets $w_n\coloneqq w_1\oplus\dots\oplus w_{n-1}\oplus w$. It now follows from the security of the MPC that the views of the parties $\{P_i\}_{i\in T}$ generated using the inputs $(w_i,r'_i)_{i\in[n]}$ for random $r'_i$ are indistinguishable from the views $\{P_i\}_{i\in T}$ generated using the inputs $(w'_i,r_i)_{i\in[n]}$ for random $w'_i$. This implies the indistinguishability of $\cH_0$ and $\cH_1$ and concludes the transition.

        \paragraph{Indistinguishability of hybrids $\cH_1$ and $\cH_2$:} The indistinguishability of the hybrids $\cH_1$ and $\cH_2$ follows directly from the fact that
        \[
            \{(w_i|i\in T)|(w_1,\dots,w_n)\gets\share(w)\}=\{(w_i|i\in T)|(w_1,\dots,w_n)\gets\share'(0)\}
        \]
        where $\share'(w)$ samples $w_1,\dots,w_{n-1}$ randomly and sets $w_n\coloneqq\bigoplus_{i\in[n-1]}w_i$. More precisely, for a challenge $w,T$ submitted by an adversary, the above equality implies that it is (perfectly) indistinguishable if the values $w_1,\dots,w_{n}$ used as the input to the MPC protocol $\Pi$ are generated using $\share'(w)$ or using $\share'(0)$. This implies that the hybrids $\cH_1$ and $\cH_2$ are perfectly indistinguishable.

        Combining both of the above analyses yields the lemma.
    \end{proof}
    Taking into account~\cref{thm:dfns}, we observe that the above defined secret sharing scheme is secure against superposition attacks as long as only $t/2$ shares are revealed.
    
    \begin{claim}
        \label[claim]{claim:h3-to-h4}
        Let $\Pi$ be a perfectly $t$-private MPC protocol, then the hybrids $\cH_3$ and $\cH_4$ are perfectly indistinguishable.
    \end{claim}

    \begin{proof}

        This claim directly follows with the observation made in~\cref{lem:t-out-of-n-classical} that the above described scheme is a $t$-out-of-$n$ secret sharing scheme. Therefore, we can rely on the results of Damg\r{a}rd et al.~\cite[Theorem 1]{DamgardFNS13} which state that a perfectly secure $t$-out-of-$n$ secret sharing scheme is secure against superposition attacks as long as only $t/2$ of the shares are revealed. This directly implies the perfect indistinguishability of the hybrids $\cH_3$ and $\cH_3$ since the interaction between the prover and the verifier in the third round corresponds to revealing $t/2$ shares that have either been generated for the witness $w$ of the relation (in the case of $\cH_3$) or the opening of the commitment $c$ (in the case of $\cH_4$). This concludes the proof of the claim.
    \end{proof}

\end{proof}

\section{Superposition-attack secure, Zero-Knowledge Argument for $\QMA$}\label{sec:zk-qma}

\subsection{Quantum Secret Sharing Schemes~\cite[Section 4]{DamgardFNS13} \& MPQC}

In this section, we introduce some preliminaries on the superposition security of quantum secret sharing schemes and multiparty quantum computation.

\subsubsection{Quantum Secret Sharing Secure against Superposition Attacks}

Our protocol relies on a quantum secret sharing scheme that is secure against superposition attacks. 

\begin{definition}[Quantum Secret Sharing]
    Let $\ket{sec}=\sum_{s\in\mathbb{S}}\alpha_s\ket{s}$ be the secret quantum state. A quantum secret sharing scheme is defined as a unitary transformation that maps each basis state $\ket{s}$ plus an ancilla of appropriate size to a state $\ket{\phi_s}=\sum_{v_1,\dots,v_n\in\bits^k}\alpha_{v_1,\dots,v_n}\ket{v_1}\dots\ket{v_n}$. The content of the $i$'th register is the share of the $i$'th party.

    A quantum secret sharing scheme is secure against adversary structure $F$, if any subset of shares corresponding to a subset $A\in F$ contains no information about the secret state, but any subset of shares $B$ where $B\notin F$ allows reconstruction of the secret. It follows from the no-cloning theorem that security can only hold if $F$ has the property that for any $B\notin F$, the complement $\overline{B}$ is in $F$. 
\end{definition}

\begin{definition}[$F$-Share Capture Attacks]
    Let $\Vec{v}=v_1,\dots,v_n$ and $\ket{\Vec{v}_A}$ be the basis state where $\ket{\Vec{v}_A}=\ket{w_1}\dots\ket{w_n}$, with $w_i=v_i$ if $i\in A$ and $w_i=\bot$ otherwise. Furthermore, we specify a unitary $U$ by the following action on basis states for shares and player subsets: $U(\ket{v_1}\dots\ket{v_n}\ket{A}\ket{\bot}^{\otimes n})=\ket{\vec{v}_{\overline{A}}}\ket{A}\ket{\vec{v}_A}$. 

    In an $F$-share capture attack, the adversary gets to specify a ``query'' state $\sum_{A\in F}\alpha_A\ket{A}\ket{\bot}^{\otimes n}$. The unitary $U$ is then applied to the secret shared state as well as the query state, i.e.\ \\ $U(\ket{\phi_s}\sum_{A\in F}\alpha_A\ket{A}\ket{\bot}^{\otimes n})$. In return, the adversary receives the last two registers (which contain the corrupted sets and the captured shares).

    We say that the scheme is secure against $F$-share capture attacks if it always holds that the state the adversary gets is independent of $\ket{s}$. By linearity, security trivially extends to superpositions over different $\ket{s}$'s.
\end{definition}

\begin{theorem}[\protect Superposition Security of Quantum Secret Sharing (Proposition 1~\cite{DamgardFNS13})]
    \label{thm:dfns-quantum}
    Any quantum secret sharing scheme that is secure for adversary structure $F$ (\cref{def:adv-str}) is also secure against $F$-share capture attacks.
\end{theorem}

In the proof of~\cref{thm:mpqc-in-the-head}, more specifically the hybrid transition of~\cref{claim:qma-h4-to-h5}, we rely on the superposition security of a quantum secret sharing scheme. In this hybrid transition, we consider the circuit-to-Hamiltonian reduction over the MPQC circuit as the quantum secret sharing scheme. Note that, strictly speaking, this procedure does not satisfy all the properties of a standard quantum secret sharing scheme, since there is no well-defined reconstruction procedure for \emph{all} subsets of parties (although there is one for a particular choice of a subset). We can nevertheless rely on~\cref{thm:dfns-quantum} since the proof of this theorem only relies on the secrecy of the quantum secret sharing scheme, which is given by the above construction, and not its reconstruction property.

\subsubsection{Multiparty Quantum Computation}

Let $\cR$ be a quantum circuit which has $n$ sets of input and output registers.
Let $\Pi$ be an $n$-party MPQC protocol computing the quantum circuit $\cR$.
$\Pi$ is executed on the following set of registers: for each party $i \in [n]$, an input register $I_i$, a work register $W_i$, a message register $M_i$ (consisting of one sub-register for each party $j\neq i$ and round), and an output register $O_i$. We refer to the set of registers of each party as that party's \emph{view}. Registers $W_i$, $M_i$, and $O_i$ are initialized to $\ket{0}$. The input registers $I_i$ contain the input to $\cR$. We represent our MPQC protocol as a quantum circuit $\cC = U_T \ldots U_1$ where the $U_i$ are two-qubit unitaries.

\begin{remark}
Without loss of generality, we assume that parties in the MPQC protocol do not receive randomness as input. This is because they can obtain such randomness by creating an auxiliary $\ket{+}$ state, and measuring it (and the latter measurement can of course be deferred to the very end).  \end{remark}

\subsection{Our Protocol}\label{sec:qmaprotocol}

At a high-level, our construction involves secret sharing a quantum witness amongst $n$ parties, constructing an $n$-party MPQC circuit which computes a $\QMA$ verifier, applying the quantum circuit to local Hamiltonian reduction to the MPQC circuit, and performing a commit-and-open style quantum protocol analogous to the classical MPC-in-the-head paradigm.

We describe our protocol in \cref{fig:zk-for-qma} and \cref{fig:zk-for-qma-protocol}. In our protocol, we draw upon the classical MPC-in-the-head technique of~\cite{IshaiKOS09} as well as the use of dual-mode commitments to achieve superposition-security from~\cite{DamgardFNS13}.

\begin{figure}[!ht]
\begin{framed}
\centering
\begin{minipage}{1.0\textwidth}
\begin{center}
    \underline{Superposition-Attack Secure Zero-Knowledge Argument for $\QMA$ from MPQC}
\end{center}

\vspace{2mm}

Let $(\Gen_H, \Gen_B, \DualCom)$ be a dual-mode commitment scheme.
Let $(\Gen, \Com)$ be a perfectly binding, computationally hiding commitment scheme with randomness of length $\lambda$.
Let $\Pi$ be an $n$-party statistically correct and perfectly $2t$-private MPQC protocol. Let $\cL = (\cL_{yes}, \cL_{no})$ be a $\QMA$ promise problem and $\cM$ be a quantum polynomial-time family of circuits $\cM = \{\cM_n\}_{n \in \bbN}$ for $\cL$ that accepts yes instances with probability $1 - \negl(n)$. Let $x \in \cL_{yes}$ with witness $\ket{w}$ of length $m$.

\vspace{0.2cm}

\noindent {$\underline{\textsc{Setup}(1^\lambda)}$}:
Let $\crs = (\ck_B, \ck, c)$ where two commitment keys and a commitment are sampled as follows: $\ck_B \gets \Gen_B(1^\lambda)$, $\ck \gets \Gen(1^\lambda)$, and $c \gets \Com(\ck, 0)$. Outputs $\crs$.

\vspace{0.2cm}

Let $\cR$ be a quantum circuit (with $(x, c, \ck)$ hardcoded) that on input $\ket{a_1, b_1, r_1} \tensor \ldots \tensor \ket{a_{n-1}, b_{n-1}, r_{n-1}} \tensor \ket{\psi, r_n}$, where $a_i,b_i \in \{0,1\}^m$ and $r_i \in \{0,1\}^{\lambda}$, does the following:
\begin{itemize}
    \item Defines $(a, b) = \xor_{i \in [n-1]} (a_i, b_i)$ and $r = \xor_{i \in [n]} r_i$.
    \item Computes $\ket{w'} = Z^b X^a \ket{\psi}$.
    \item If $\cM(x, \ket{w'}) = 1$ or $c \equiv \Com(\ck, 1; r)$, outputs $1$. Otherwise, outputs $0$.
\end{itemize}

Let $\cC$ be a quantum circuit for the MPQC protocol $\Pi$ computing the quantum circuit $\cR$.
Apply the compiler described in \cref{sec:history}, with details from~\cref{sec:mpqc-to-cth}, to $\cC$. This defines a Hamiltonian $H$.
Let $\ell$ be the number of qubits in the work, message, and output registers.

\vspace{0.3cm}

Repeat \cref{fig:zk-for-qma-protocol} $\poly(\lambda)$ times in parallel, each on a separate copy of the witness. The output of the verifier is defined as the majority vote over all the instances. 

\end{minipage}
\end{framed}
\caption{MPQC in the Head (Part I)}
\label{fig:zk-for-qma}
\end{figure}

\begin{figure}[!ht]
\begin{framed}
\centering
\begin{minipage}{1.0\textwidth}

\noindent {$\underline{\textsc{Prover}_1(\crs, x, \ket{w})}$}:
\begin{itemize}
    \item Sample $(a, b)$ uniformly at random from $\zo^m \times \zo^m$. Sample $n-1$ shares $(a_1, b_1), \ldots, (a_{n-1}, b_{n-1}) \in \zo^m \times \zo^m$ such that $\xor_{i \in [n-1]} (a_i, b_i) = (a, b)$. Compute a quantum OTP on the witness, $\ket{\psi} = X^aZ^b \ket{w}$. Sample $r_i$ uniformly at random from $\{0, 1\}^\lambda$ for $i \in [n]$.
    
    \item Let $\ket{\phi_{hist}}$ be the history state corresponding to the circuit $\cC$ defined in~\cref{fig:zk-for-qma} 
on input $\ket{a_1, b_1, r_1} \tensor \ldots \tensor \ket{a_{n-1}, b_{n-1}, r_{n-1}} \tensor \ket{\psi, r_n}$.

    \item Let $m'$ be the number of qubits in $\ket{\phi_{hist}}$. Sample $(a', b')$ uniformly at random from $\zo^{m'} \times \zo^{m'}$. Compute a quantum OTP on the history state, $\ket{\psi_{hist}} = X^{a'}Z^{b'} \ket{\phi_{hist}}$.
    
    \item Compute commitments $\alpha_i = \DualCom(\ck_B, (a_i', b_i'); r_i')$ where $a_i'$ and $b_i'$ are the keys corresponding to the $i$th party's registers and $r_i'$ is sampled uniformly at random for $i \in [n]$. Define openings $\zeta_i = ((a_i', b_i'), r_i')$ for $i \in [n]$.

    \item Output $(\alpha = (\ket{\psi_{hist}}, \alpha_1, \ldots, \alpha_n), \zeta = (\zeta_1, \ldots, \zeta_n))$.
\end{itemize}

\noindent {$\underline{\textsc{Verifier}_1(\crs, x, \alpha)}$}:
Sample $S$ according to $p_S$ as outlined in \cref{sec:history}.
Let $\beta = (\beta_1, \beta_2)$ specify a set of $2$ parties which are sampled uniformly at random conditioned on opening the two qubits which are measured by $S$.
Output $(\beta, \xi = S)$.

\vspace{0.3cm}

\noindent {$\underline{\textsc{Prover}_2(\crs, x, \alpha, \beta, \zeta)}$}:
Output $\gamma = \{\zeta_{\beta_1}, \zeta_{\beta_2}\}$.

\vspace{0.3cm}

\noindent {$\underline{\textsc{Verifier}_2(\crs, x, \alpha, \beta, \gamma, \xi)}$}:
Output $1$ if and only if the following checks pass:
\begin{itemize}
    \item The openings are valid $\alpha_{\beta_i} \equiv \DualCom(\ck_B, (a_{\beta_i}', b_{\beta_i}'); r_{\beta_i}')$
for $i \in [2]$.
    \item Apply $Z^{b_{\beta_i}'}X^{a_{\beta_i}'}$ to the $\beta_i^{th}$ qubit in $\ket{\psi_{hist}}$ for all $i \in [2]$ to get a resulting state $\ket{\psi_{hist}'}$.
The check in \cref{sec:history} with respect to the previously chosen measurement $S$ passes for $\ket{\psi_{hist}'}$ at indices $\{\beta_i\}_{i \in [2]}$.
\end{itemize}
Else, output $0$.

\vspace{0.1cm}

\end{minipage}
\end{framed}
\caption{MPQC in the Head (Part II)}
\label{fig:zk-for-qma-protocol}
\end{figure}

\subsection{Analysis}

\begin{theorem}
    \label{thm:mpqc-in-the-head}
    Let $n \in \bbN$ and $t < n/3$.
    Let a dual-mode commitment $(\Gen_H, \Gen_B, \DualCom)$ (\cref{def:dm-com}), a perfectly binding commitment $(\Gen, \Com)$ (\cref{def:com}), and a $n$-party, statistically correct and perfectly $2t$-private MPQC protocol $\Pi$ be given (\cref{def:mpqc}).

    The protocol in \cref{fig:zk-for-qma,fig:zk-for-qma-protocol} is a statistically sound, computationally zero-knowledge against superposition-attacks for $\QMA$ in the common reference string model.
\end{theorem}

\begin{proof}    
As we will see, in order to prove \cref{thm:mpqc-in-the-head}, it suffices to provide a proof with respect to a single repetition of the protocol in \cref{fig:zk-for-qma,fig:zk-for-qma-protocol}.
    
    We will first show that it suffices to analyze completeness and soundness for one parallel repetition. In what follows will prove that a single repetition has a noticeable completeness-soundness gap, in other words:
    \[
    p^{acc}_{yes} - p^{acc}_{no} \geq \frac{1}{q(n)}
    \]
    for a polynomial $q(n)$ with $n\in\bbN$.    
One can reduce the completeness and the soundness error using a standard technique: Composing the protocol in parallel with multiple copies of the witness and taking the majority vote. Hence, while our protocol necessarily uses multiple repetitions to achieve \cref{thm:mpqc-in-the-head}, it suffices to analyze completeness and soundness for one parallel repetition.
    
    We now argue that it also suffices to analyze zero-knowledge for one parallel repetition. Our protocol, without considering the trapdoor in the CRS, is a witness-indistinguishable proof, and witness indistinguishability composes in parallel. The simulation for the parallel-repeated version of the protocol can be then argued using the technique introduced in~\cite{FeigeLS90}, where the real witness is gradually substituted by the trapdoor in the CRS, one repetition at a time.

    \paragraph{Statistical Completeness.} Let $\cL = (\cL_{yes}, \cL_{no}) \in \QMA$ be an input promise problem, $\cA$ be an arbitrary, unbounded-size quantum circuit, and $x$ be an arbitrary instance $x \in \cL_{yes}$ with corresponding witness $\ket{w}$. Let $\ket{w'} = \ket{a_1, b_1, r_1} \tensor \ldots \tensor \ket{a_{n-1}, b_{n-1}, r_{n-1}} \tensor \ket{\psi, r_n}$ be the honest prover's input.

    We proceed by looking at the honest verifier's checks when interacting with an honest prover. 
    Since the dual-mode commitment is deterministic when its randomness is provided, the honest verifier's first check that the openings are valid will pass with probability $1$. 
    By the perfect correctness of the quantum OTP, the reduced density of $\ket{\psi_{hist}'}$ on registers $\beta_1$ and $\beta_2$ is identical to that of $\ket{\phi_{hist}}$ (on the same registers). Furthermore, by the statistical correctness of the MPQC $\Pi$, the protocol $\Pi$ will accept with probability $1 - \negl(\lambda)$. Overall, this implies that the Hamiltonian $H$ has a ground state with a negligible eigenvalue, and furthermore the ground state is exactly $\ket{\phi_{hist}}$. Thus, verification will succeed probability negligibly close to $1/2$.

    \paragraph{Statistical Soundness.}
    Since $x \in \cL_{no}$, then for all quantum states , the probability that $\cM$ accepts a given state is bounded away from $1$, by at least some inverse polynomial term $1/p(n)$. Furthermore, since the commitment $c$ is perfectly binding, the same bound must hold for the circuit $\cR$. By the statistical correctness of $\Pi$, we can also conclude that an honest execution of $\cC$ would reject any input state with probability at least $1/p(n)$. Consequently, the smallest eigenvalue of the Hamiltonian $H$ is at least $1/q(n)$, for some polynomial $q(n)$. 
    
    Let us now consider the commitment message $\alpha$ as computed by $\cA$. This commitment is a quantum OTP implemented with a perfectly binding dual-mode commitment. Since it is perfectly binding, there is a well-defined unique state $\rho$ contained in $\alpha$. 
    It follows that the state must be independent of the random choices of the verifier, and consequently, the verifier of the overall protocol accept with probability at most $1/2 - 1/\poly(n)$.

    \paragraph{Superposition-Secure Computational Zero-Knowledge.}
    As in the analysis for the NP case, to model superposition access of the adversary, we consider a classical algorithm $\sP'_2$ that has values $(\crs, x, \alpha, \zeta)$ hardcoded and takes as an input only $\beta$ and then, internally, runs $\sP_2(\crs,\allowbreak x, \alpha, \beta, \zeta)$. Given the classical algorithm for $\sP'_2$, there also exists a unitary $U_{\sP'_2}$ implementing $\sP'_2$ quantumly. When the verifier now submits a superposition query $\sum_\iota \ket{\beta_\iota}$, it is possible to compute the reply, an opening over all the challenges contained in this superposition, by applying the unitary $U_{\sP'_2}$ controlled on the challenge $\sum_\iota \ket{\beta_\iota}$. More precisely, we apply \[U_{\sP'_2}\sum_\iota \ket{\beta_\iota}\ket{0}_\regM=\sum_\iota \ket{\beta_\iota}\ket{\gamma_\iota}_\regM\]
    and then return the register $\regM$ as a reply to the verifier.
    
    We define a two-part simulator $\Sim=(\Sim_1, \Sim_2)$. The first algorithm $\Sim_1$ takes as input an instance $x$:
    \begin{itemize}
        \item Computes $\crs = (\ck_H, \ck, c)$ by sampling two commitment keys $\ck_H \gets \Gen_H(1^\lambda)$, $\ck \gets \Gen(1^\lambda)$, and computing the commitment $c = \Com(\ck, 1; r)$ with uniform randomness $r$.

        \item Samples $a, b$ uniformly at random from $\zo^m$ and compute a quantum OTP encryption of the $\ket{0}$ state, $\ket{\psi} = X^aZ^b \ket{0}$.
        
        \item Samples $r_i$ uniformly at random from $\{0, 1\}^\lambda$ for $i \in [n-1]$ and set $r_n=r\xor_{i\in[n-1]}r_{i}$. Sets $x' = (x, c, \ck)$.

        \item Defines the history state $\ket{\phi_{hist}}$ as described in \cref{sec:history} to the MPQC protocol $\Pi$ computing the quantum circuit $\cR$ given input $\ket{a_1, b_1, r_1} \tensor \ldots \tensor \ket{a_{n-1}, b_{n-1}, r_{n-1}} \tensor \ket{\psi, r_n}$ with $a_i,b_i$ randomly sampled for all $i\in[n-1]$. \footnote{Here, the circuit $\cR$ will decrypt the state $\ket{\psi}$ by xoring the random values $a_i,b_i$ to obtain $a$ and $b$ which will lead to a wrong witness and therefore the xor of the opening shares $r_1,\dots,r_n$ are used as the witness.}.

        \item Let $m'$ be the number of qubits in $\ket{\psi_{hist}}$. Sample $a', b'$ uniformly at random from $\zo^{m'}$. Compute a quantum OTP on the history state, $\ket{\psi_{hist}} = X^{a'}Z^{b'} \ket{\phi_{hist}}$.
        
        \item Compute commitments $\alpha_i = \DualCom(\ck_B, (a_i', b_i'); r_i')$ where $a_i'$ is the $i$th bit of $a'$ (similarly for $b_i'$) with uniformly random $r_i'$ for $i \in [m']$. Define openings $\zeta_i = ((a_i', b_i'), r_i')$ for $i \in [n]$.
        
        \item Outputs $(\crs, \alpha = (\alpha_1, \ldots, \alpha_n), \zeta = (\zeta_1, \ldots, \zeta_n))$.
    \end{itemize}
    The second algorithm $\Sim_2$ takes input $(\crs, x, w, \alpha, \sum_\iota \ket{\beta_\iota}, \zeta)$ as input where $\sum_\iota \ket{\beta_\iota}$ may be a superposition query, hardcodes $(\crs, x, w, \alpha, \zeta)$ into $U_{\sP_2'}$, and outputs $U_{\sP_2'}(\sum_\iota \ket{\beta_\iota})$.
 
    Let a relation $\cR$ and a quantum polynomial-size quantum circuit $\cA$ that can send superposition queries.
    We define the following series of hybrids, for any $\lambda \in \bbN$ and instance-witness pair $(x, w) \in \cR$ where $|x| = \lambda$, which we will argue are all indistinguishable:
    \begin{itemize}
        \item $\cH_0$: Same as the real world. 
        
        $\crs \gets \Setup(1^\lambda)$, $(\alpha, \zeta) \gets \sP_1(\crs, x, w)$, $(\sum_\iota \ket{\beta_\iota}, \xi) \gets \cA(\crs, x, \alpha)$, hardcode $U_{\sP_2'}$ with $(\crs, x, w, \alpha, \zeta)$, compute $U_{\sP'_2}(\sum_\iota \ket{\beta_\iota}\ket{0}_\regM)=\sum_\iota \ket{\beta_\iota}\ket{\gamma_\iota}_\regM$, and where register $\regM$ is output.

        \item $\cH_1$: 
        Same as $\cH_0$ except the commitment in the $\crs$ is a commitment to $1$. 
        
        $\ck_B \gets \Gen_B(1^\lambda)$, $\ck \gets \Gen(1^\lambda)$, $c \gets \Com(\ck, 1)$, $\crs = (\ck_B, \ck, c)$, $(\alpha, \zeta) \gets \sP_1(\crs, x, w)$, $(\beta, \xi) \gets \cA(\crs, x, \alpha)$, hardcode $U_{\sP_2'}$ with $(\crs, x, w, \alpha, \zeta)$, compute $U_{\sP'_2}(\sum_\iota \ket{\beta_\iota}\ket{0}_\regM)=\sum_\iota \ket{\beta_\iota}\ket{\gamma_\iota}_\regM$, and where register $\regM$ is output.
        
        The indistinguishability between this and the previous hybrid follows as in the NP case, i.e., from the hiding of the commitment scheme $\Com$. Therefore, we refer to the proof of \cref{claim:h0-to-h1} to argue the indistinguishability between $\cH_0$ and $\cH_1$.

        \item $\cH_2$: 
        Same as $\cH_1$ except the dual mode commitment is in hiding mode. 
        
        $\ck_H \gets \Gen_H(1^\lambda)$, $\ck \gets \Gen(1^\lambda)$, $c \gets \Com(\ck, 1)$, $\crs = (\ck_H, \ck, c)$, $(\alpha, \zeta) \gets \sP_1(\crs, x, w)$, $(\sum_\iota \ket{\beta_\iota}, \xi) \gets \cA(\crs, x, \alpha)$, hardcode $U_{\sP_2'}$ with $(\crs, x, w, \alpha, \zeta)$, compute $U_{\sP'_2}(\sum_\iota \ket{\beta_\iota}\ket{0}_\regM)=\sum_\iota \ket{\beta_\iota}\ket{\gamma_\iota}_\regM$, and where register $\regM$ is output.

        The indistinguishability between this and the previous hybrid follows as in the NP case, i.e., from the indistinguishability of modes from the dual-mode commitment scheme. Therefore, we refer to the proof of \cref{claim:h0-to-h1} to argue the indistinguishability between $\cH_1$ and $\cH_2$.
         
        \item $\cH_3$: It works the same as $\cH_2$ except that the dual commitments in the prover's first message are replaced with commitments to $0$, and the binding of the commitments are broken to open them to the prover's third message.  This hybrid runs in exponential time.

        For this hybrid, and the following, we define the following polynomial-size classical circuit $\cB$ which is hardcoded with $\{((a'_i, b'_i), r'_i)\}_{i \in [n]}$: output $\{((a'_{\beta_i}, b'_{\beta_i}), r'_{\beta_i})\}_{i \in [t]}$.
        This classical circuit can be implemented as a unitary $U_{\cB}$.
        
        $\ck_H \gets \Gen_H(1^\lambda)$, $\ck \gets \Gen(1^\lambda)$, $c \gets \Com(\ck, 1)$, $\crs = (\ck_H, \ck, c)$, $\alpha = (\ket{\psi_{hist}}, \alpha_1, \ldots, \alpha_n)$ where $\alpha_i = \DualCom(\ck_H, (0,0))$ and $\ket{\psi_{hist}}$ is generated as described in $\sP_1$, $(\sum_\iota \ket{\beta_\iota}, \xi) \gets \cA(\crs, x, \alpha)$, computes valid openings $(r'_1, \ldots, r'_n)$ to $((a'_1,b'_1), \ldots, (a'_n,b'_n))$ (by running in unbounded-time), hardcodes $U_{\cB}$ with $\{((a'_i, b'_i), r'_i)\}_{i \in [n]}$, and computes $U_{\cB}(\sum_\iota \ket{\beta_\iota}\ket{0}_\regM)=\sum_\iota \ket{\beta_\iota}\ket{\gamma_\iota}_\regM$ where $\regM$ is output to the adversary.
        
        The indistinguishability between this and the previous hybrid follows from the statistical hiding property of the dual-mode commitment scheme. 
        We prove the indistinguishability of $\cH_2$ and $\cH_3$ in \cref{claim:h2-to-h3}.

        \item $\cH_4$: Same as $\cH_3$ except that the messages of the MPQC protocol are teleported into the one-time pad encryptions and the binding of the commitments are broken to open them to the corresponding teleportation keys. This hybrid also runs in exponential time.

        In this, and the following hybrids, we consider a modified (quantum) circuit $U_{\cB'}$. $U_{\cB'}$ also works controlled on the challenge of the adversary but, before outputting the reply to the query, it executes the teleportation procedure for all the indices contained in the challenge query of the adversary and then uses the output bits of the teleportation procedure to brute force the opening as in the procedure $\cB$. More formally, $U_{\cB'}$ takes as an input the state $\sum_\iota \ket{\beta_\iota}\ket{\phi_{hist}}\ket{\epr}_A\ket{0}_\regT\ket{0}_\regM$ then applies the teleportation unitary $U_T$ controlled with $\sum_\iota \ket{\beta_\iota}$ on the corresponding registers of $\ket{\epr}_A\ket{0}_\regT$. For the classical teleportation bits, contained in the register $\regT$, are then used as the values for which the openings for the dual-mode commitments are (coherently) brute forced\footnote{Here, we assume that the classical dual-mode commitments are hardcoded inside the unitary}. The output of this brute force operation are then contained in the register $\regM$, which is then output to the adversary.
        
        $\ck_H \gets \Gen_H(1^\lambda)$, $\ck \gets \Gen(1^\lambda)$, $c \gets \Com(\ck, 1)$, $\crs = (\ck_H, \ck, c)$, sample $n$ EPR pairs $\ket{\epr}_{A,B}\coloneqq\ket{\epr}_{A,B,1},\dots,\ket{\epr}_{A,B,n}$, $\alpha = (\ket{\epr}_{B}, \alpha_1, \ldots, \alpha_n)$ where $\alpha_i = \DualCom(\ck_H,\allowbreak (0,0))$, $(\sum_\iota \ket{\beta_\iota}, \xi) \gets \cA(\crs, x, \alpha)$, apply the unitary $U_{\cB'}(\sum_\iota \ket{\beta_\iota}\allowbreak\ket{\phi_{hist}}\ket{\epr}_A\ket{0}_\regT\ket{0}_\regM)$ (by running in unbounded-time) with $\ket{\phi_{hist}}$ being generated as described in $\sP_1$, and output $\regM$ to the adversary.
        
        The indistinguishability between this and the previous hybrid follows from the (perfect) security of the quantum one-time pad. We prove the indistinguishability of $\cH_3$ and $\cH_4$ in \cref{claim:qma-h3-to-h4}.
  
        \item $\cH_5$: Same as $\cH_4$ except that the first message computed by the prover is completely independent of the witness $\ket{w}$ and computed for the value $\ket{0}$ instead. This also implies that the history state $\ket{\phi_{hist}}$, is generated using the shares $r_1,\dots,r_n$ of the opening $r$ from the commitment $c$ instead of the decryption keys $(a,b)$ for the quantum one-time pad ciphertext. This hybrid also runs in exponential time.

        See the definition of $U_{\cB'}$ in the previous hybrid.
        
        $\ck_H \gets \Gen_H(1^\lambda)$, $\ck \gets \Gen(1^\lambda)$, $c \gets \Com(\ck, 1)$, $\crs = (\ck_H, \ck, c)$, generate $\ket{\psi}$ by encrypting the $\ket{0}$ state, i.e., $\ket{\psi}=X^aZ^b\ket{0}$ for random $a,b$, generate $\ket{\phi_{hist}}$ using the secret shares $r_1,\dots,r_n$ of the opening $r$ of the commitment $c$ and using random $a_i,b_i$ for all $i\in[n]$ as well as $\ket{\psi}$, sample $n$ EPR pairs $\ket{\epr}_{A,B}\coloneqq\ket{\epr}_{A,B,1},\dots,\ket{\epr}_{A,B,n}$, $\alpha = (\ket{\epr}_{B}, \alpha_1, \ldots, \alpha_n)$ where $\alpha_i = \DualCom(\ck_H, (0,0))$, $(\sum_\iota \ket{\beta_\iota}, \xi) \gets \cA(\crs, x, \alpha)$, apply the unitary $U_{\cB'}(\sum_\iota \ket{\beta_\iota}\allowbreak\ket{\phi_{hist}}\ket{\epr}_A\ket{0}_\regT\ket{0}_\regM)$ (by running in unbounded-time), and output $\regM$ to the adversary.
        
        The indistinguishability between this and the previous hybrid follows from the (perfect) security of the one-time pads and the (perfect) $2t$-privacy of the underlying MPC protocol.  We prove the indistinguishability of $\cH_4$ and $\cH_5$ in \cref{claim:qma-h4-to-h5} using the results of~\cref{lem:t-out-of-n-quantum} where we abstract the computation of the first message of the prover as a quantum secret sharing scheme.

        \item $\cH_6$: Same as $\cH_6$ except that the EPR pairs are, again, replaced with quantum one-time pad encryptions of the history state $\ket{\phi_{hist}}$. In this hybrid, it is also possible to, again, rely on the unitary $U_{\cB}$ to program the opening of the dual-mode commitments. In more detail:
        
        $\ck_H \gets \Gen_H(1^\lambda)$, $\ck \gets \Gen(1^\lambda)$, $c \gets \Com(\ck, 1)$, $\crs = (\ck_H, \ck, c)$, generate $\ket{\psi}$ by encrypting the $\ket{0}$ state, i.e., $\ket{\psi}=X^aZ^b\ket{0}$ for random $a,b$, generate $\ket{\phi_{hist}}$ using the secret shares $r_1,\dots,r_n$ of the opening $r$ of the commitment $c$ and using random $a_i,b_i$ for all $i\in[n]$ as well as $\ket{\psi}$, $\alpha = (\ket{\psi_{hist}}, \alpha_1, \ldots, \alpha_n)$ where $\alpha_i = \DualCom(\ck_H, (0,0))$ and $\ket{\psi_{hist}}$ is generated as described in $\textsc{Prover}_1$ using $\ket{\phi_{hist}}$ as generated above, $(\sum_\iota \ket{\beta_\iota}, \xi) \gets \cA(\crs, x, \alpha)$, computes valid openings $(r'_1, \ldots, r'_n)$ to $((a'_1,b'_1), \ldots, (a'_n,b'_n))$ (by running in unbounded-time), hardcodes $U_{\cB}$ with $\{((a'_i, b'_i), r'_i)\}_{i \in [n]}$, and computes $U_{\cB}(\sum_\iota \ket{\beta_\iota}\ket{0}_\regM)=\sum_\iota \ket{\beta_\iota}\ket{\gamma_\iota}_\regM$ where $\regM$ is output to the adversary.
        
        The indistinguishability between this and the previous hybrid follows from the (perfect) security of the quantum one-time pad. We prove the indistinguishability of $\cH_3$ and $\cH_4$ in \cref{claim:qma-h3-to-h4}.

        \item $\cH_7$: Same as $\cH_6$ except that the dual-mode commitments are, again, generated using the actual one-time pad keys $(a',b')$, which are used to encrypt the history state $\ket{\phi_{hist}}$.

        $\ck_H \gets \Gen_H(1^\lambda)$, $\ck \gets \Gen(1^\lambda)$, $c \gets \Com(\ck, 1)$, $\crs = (\ck_H, \ck, c)$, generate $\ket{\psi}$ by encrypting the $\ket{0}$ state, i.e., $\ket{\psi}=X^aZ^b\ket{0}$ for random $a,b$, generate $\ket{\phi_{hist}}$ using the secret shares $r_1,\dots,r_n$ of the opening $r$ of the commitment $c$ and using random $a_i,b_i$ for all $i\in[n]$ as well as $\ket{\psi}$, $\alpha = (\ket{\psi_{hist}}, \alpha_1, \ldots, \alpha_n)$ where $\alpha_i = \DualCom(\ck_H, (a'_i,b'_i);r_i'),\zeta=\{\zeta_i=((a'_i,b'_i),r'_i)\}_{i\in[n]}$ and $\ket{\psi_{hist}}$ is generated as described in $\textsc{Prover}_1$ using $\ket{\phi_{hist}}$ as generated above, $(\sum_\iota \ket{\beta_\iota}, \xi) \gets \cA(\crs, x, \alpha)$, hardcodes $U_{\sP_2'}$ with $(\crs, x, \alpha, \zeta)$, compute $U_{\sP'_2}(\sum_\iota \ket{\beta_\iota}\ket{0}_\regM)=\sum_\iota \ket{\beta_\iota}\ket{\gamma_\iota}_\regM$, and output register $\regM$ to the adversary.
        
        The statistical indistinguishability between this and the previous hybrid follows from the statistical security of the dual-mode commitments. The proof proceeds analogously to the proof of the indistinguishability between $\cH_6$ and $\cH_7$ (\cref{claim:h2-to-h3}).

\end{itemize}

    \begin{claim}
        \label[claim]{claim:qma-h3-to-h4}
        The hybrids $\cH_3$ and $\cH_4$ are perfectly indistinguishable.
    \end{claim}
    \begin{proof}
        This claim follows directly using the observation that one half of an EPR pair is perfectly indistinguishable from a quantum one-time pad encryption. Therefore, sampling $\ket{\psi_{hist}}$ as half of an EPR pair is indistinguishable to providing an actual encryption of the history state $\ket{\phi_{hist}}$ as the state $\ket{\psi_{hist}}$.

        Now, when the adversary asks its challenge, it follows that, after the teleportation protocol has been executed, that the classical bits $a,b$ obtained from the teleportation procedure allow to reconstruct the state state $\ket{\phi_{hist}}$ from the state $\ket{\psi_{hist}}$. This happens using the same operations as in the case that the state $\ket{\psi_{hist}}$ corresponds to a one-time pad encryption. In the next step, the reduction runs in unbounded time to compute the openings for the challenged dual-mode commitments that corresponds to the bits $a,b$ obtained from the quantum teleportation protocol. This concludes the proof of the claim. 
    \end{proof}

    Before proceeding with the proof of~\cref{claim:qma-h4-to-h5}, we cast the first three steps of the prove procedure $\textsc{Prover}_1(\crs,x,\ket{w})$, together with some parts of the CRS, as the sharing procedure of a secret sharing scheme for a secret $\ket{w}$ of length $m$. More precisely:
    \begin{description}
        \item[$\Setup(1^\secpar)$:] Generate a commitment key $\ck \gets \Gen(1^\lambda)$ for a perfectly binding, computationally hiding commitment scheme and a commitment $c \gets \Com(\ck, 1)$. Output $\crs = (\ck, c)$.
        
        \item[$\share(\crs,\ket{w},n)$:] Sample $(a_1, b_1),\dots,(a_{n-1},b_{n-1})$ uniformly at random from $\zo^m \times \zo^m$. Set $(a,b)\coloneqq\bigoplus_{i \in [n-1]} (a_i, b_i)$. Compute a quantum OTP of the witness, $\ket{\psi} = X^aZ^b \ket{w}$ and sample $r_i$ uniformly at random from $\{0, 1\}^\lambda$ for $i \in [n]$. Compute the history state $\ket{\phi_{hist}}$ corresponding to the circuit $\cC$ defined in~\cref{fig:zk-for-qma} on input $\ket{a_1, b_1, r_1} \tensor \ldots \tensor \ket{a_{n-1}, b_{n-1}, r_{n-1}} \tensor \ket{\psi, r_n}$ and using $\crs$. Let $m'$ be the number of qubits in $\ket{\phi_{hist}}$. Sample $(a', b')$ uniformly at random from $\zo^{m'} \times \zo^{m'}$. Compute a quantum OTP of the history state, $\ket{\psi_{hist}} = X^{a'}Z^{b'} \ket{\phi_{hist}}$. Output $((a_i',b_i')_{i\in[n]},\ket{\psi_{hist}})$ with $a_i'$ and $b_i'$ being the keys corresponding to the $i$th party's registers. $(a_i',b_i')$ is the share of the $i$th party and $\ket{\psi_{hist}}$ is made public.

        \item[$\rec(\crs,\ket{\psi_{hist}},(a_i',b_i')_{i\in[n]})$:] Compute $ \ket{\phi_{hist}}=Z^{b'}X^{a'}\ket{\psi_{hist}}$. In the next step, undo all of the controlled $U_t,\dots,U_1$ of the history state $ \ket{\phi_{hist}}$ using $\crs$ to obtain the inputs $\ket{a_1, b_1, r_1} \tensor \ldots \tensor \ket{a_{n-1}, b_{n-1}, r_{n-1}} \tensor \ket{\psi, r_n}$ of the circuit $\cC$. Next, compute and output $\ket{w}=Z^bX^a\ket{\psi}$.
    \end{description}
	
	In the next step, we prove that the presented secret sharing scheme achieves perfect $t'$-out-of-$n$ security.
	
	\begin{claim}
		\label[claim]{lem:t-out-of-n-quantum}
		Let $\Pi$ be a perfectly $t'$-private MPQC protocol, then the above secret sharing scheme is perfectly secure against classical $G$ attacks where $G$ contains all subsets of $[n]$ of size $t'$.
	\end{claim}
	\begin{proof}
		To prove this lemma, it suffices to prove that the above described quantum secret sharing scheme is a $t'$-out-of-$n$ secret sharing scheme, following~\cref{rem:sss}. We prove this using the following hybrids:
		\begin{description}
			\item[Hybrid $\cH_0$:] This hybrid corresponds to the case where (above) the secret sharing scheme is executed using the secret $\ket{w}$.			

            \item[Hybrid $\cH_1$:] In this hybrid, the positions of the history state that the adversary does not request as shares are replaced with encryptions of $\ket{0}$. In more detail, if the adversary requests the set of shares $T$ then the state that it will receive is the state $\ket{\psi_{hist}} = X^{a'_1}Z^{b'_1} \ket{\phi_{hist,1}} \tensor \dots \tensor X^{a'_1}Z^{b'_1} \ket{\phi_{hist,n'}}$ where $\ket{\phi_{hist,i}}=\ket{0}$ for all $i\in T$. The perfect indistinguishability between this and the previous hybrid follows from the security of the quantum one-time pad.
   
            \item[Hybrid $\cH_2$:] This hybrid corresponds to the setting where the messages of the MPQC protocol $\Pi$ are generated using the opening of the commitment $c$ as an input instead of the inputs $\ket{a_1, b_1, r_1} \tensor \ldots \tensor \ket{a_{n-1}, b_{n-1}, r_{n-1}} \tensor \ket{\psi, r_n}$ where $(a_1,b_1),r_1,\dots,(a_{n-1},b_{n-1}),\allowbreak r_{n-1},r_n$ are randomly sampled $(a,b)\coloneqq\bigoplus_{i \in [n-1]} (a_i, b_i)$ and $\ket{\psi} = X^aZ^b \ket{w}$. The perfect indistinguishability between this and the previous hybrid follows from the perfect $t'$-privacy of the MPQC protocol $\Pi$.
			
	      	\item[Hybrid $\cH_3$:] In this hybrid, the state $\ket{\psi}$ is generated as an encryption of $\ket{0}$ instead of an encryption of $\ket{w}$. This hybrid is perfectly indistinguishable from the previous hybrid due to the secruity of the quantum one-time pad.

            Since this hybrid does not make any use of the $\ket{w}$, this hybrid concludes the security analysis.
			
\end{description}

        \paragraph{Indistinguishability of hybrids $\cH_0$ and $\cH_1$:} The (perfect) indistinguishability of the hybrids $\cH_0$ and $\cH_1$ follows from the (perfect) security of the quantum one-time pad. We construct a reduction that interacts with the challenger of the quantum one-time pad and simulates the hybrid $\cH_0$ or $\cH_1$ to the adversary. Here, the underlying challenger runs $T$ different quantum OTP instances, all using the same challenge bit. Such a challenger directly follows from the standard quantum OTP challenger. In the first step, the reduction sets up the $\crs$ and, afterwards, receives the values $(T,\ket{w})$ from the adversary. Then, the reduction proceeds with the execution of the sharing procedure until it comes to the encryption of the history state $\ket{\phi_{hist}}$. Here, the reduction submits $(\ket{\phi_{hist,i}},\ket{0})_{i\in[T]}$ to the underlying challenger and obtains as a reply the ciphertexts $\{\ket{\psi_{hist,i}}\}_{i\in T}$. The remaining encryptions, i.e., $\ket{\psi_{hist,i}}=X^{a'_i}Z^{b'_i}\ket{\phi_{hist,i}}$ with random $(a'_i,b'_i)$ for all $i\in[n']\setminus T$, are generated as described in the sharing procedure. The reduction then sets $\ket{\psi_{hist}}\coloneqq\ket{\psi_{hist,1}}\tensor\dots\tensor\ket{\psi_{hist,n'}}$ and sends this state to the adversary. Our reduction will preserve the distinguishing advantage of $\cD$ and we observe that if the underlying challenger encrypts the actual history state the reduction simulates hybrid $\cH_0$ and if it encrypts $\ket{0}$ it simulates $\cH_1$. This results in the perfect indistinguishability of $\cH_0$ and $\cH_1$.
        
		\paragraph{Indistinguishability of hybrids $\cH_1$ and $\cH_2$:} We prove the indistinguishability of the hybrids $\cH_1$ and $\cH_2$ by constructing an adversary $\cB$ that breaks the perfect $t'$-privacy of the MPQC protocol. In the first step, the adversary $\cB$ generates the setup for the secret sharing scheme and sends $\crs$ to $\cA$. Afterwards, it receives the secret $\ket{w}$ as well as a set of indices $T$ from the adversary $\cA$. In the next step, the adversary $\cB$ samples $n-1$ shares $(a_1, b_1), \ldots, (a_{n-1}, b_{n-1}) \in \zo^m$, sets $(a, b)\coloneqq\xor_{i \in [n-1]} (a_i, b_i)$ and computes a quantum OTP encryption of the witness, $\ket{\psi} = X^aZ^b \ket{w}$. It also computes an additive secret sharing $r_1,\dots,r_n$ for the opening of the commitment $c$. Now, by applying~\cref{lem:mpqc-to-cth}, we can argue that a history state generated using the inputs that contain the key of the quantum OTP $\ket{a_1, b_1, r'_1} \tensor \ldots \tensor \ket{a_{n-1}, b_{n-1}, r'_{n-1}} \tensor \ket{\psi, r'_n}$, for the circuit induced by $\Pi_{\mathcal{C}}$ using random $r'_i$, is indistinguishable from a history state generated using the inputs that contain the trapdoor $\ket{a'_1, b'_1, r_1} \tensor \ldots \tensor \ket{a'_{n-1}, b'_{n-1}, r_{n-1}} \tensor \ket{\psi, r_n}$, for the circuit induced by $\Pi_{\mathcal{C}}$ using random $a'_i,b'_i$. This is true since both of these inputs lead to $1$ as an output of the MPQC evaluation. Since~\cref{lem:mpqc-to-cth} relies on the perfect $t$-privacy of the MPQC, it follows that the hybrids $\cH_1$ and $\cH_2$ are perfectly indistinguishable based on the $t$-privacy of the MPQC.

        \paragraph{Indistinguishability of hybrids $\cH_2$ and $\cH_3$:} The (perfect) indistinguishability of the hybrids $\cH_2$ and $\cH_3$ follows from the (perfect) security of the quantum one-time pad. We construct a reduction that interacts with the challenger of the quantum one-time pad and simulates the hybrid $\cH_2$ or $\cH_3$ to the adversary. In the first step, the reduction sets up the $\crs$, receives the values $(T,\ket{w})$ from the adversary, submits the two states $(\ket{0},\ket{w})$ to the underlying challenger and obtains as a reply the ciphertext $\ket{\psi}$. Afterwards, the reduction proceeds as described in the definition of $\cH_2/\cH_3$. We observe that none of these steps requires the decryption of $\ket{\psi}$. This is due to the fact that the history state $\ket{\phi_{hist}}$ is generated using the secret shares $r_1,\dots,r_n$ of the opening $r$ of the commitment $c$ and using random $a_i,b_i$ for all $i\in[n]$ as well as $\ket{\psi}$. Here, the secret shares $r_1,\dots,r_n$ of the opening $r$ of the commitment $c$ will be used as a witness in the circuit $\cR$. Therefore, it is not required to provide valid decryption keys for $\ket{\psi}$ to a state $\ket{w}$ that leads to $\cM(x,\ket{w})=1$. Our reduction will preserve the distinguishing advantage of $\cD$. However, this contradicts the perfect indistinguishability of the quantum one-time pad. As such, we must have that $\cH_2$ and $\cH_3$ are perfectly indistinguishable.

		Combining the above analyses yields the lemma.
	\end{proof}
	Taking into account~\cref{thm:dfns-quantum}, we observe that the above defined secret sharing scheme is secure against superposition attacks as long as only $t/2$ shares are revealed for $t'=2t$.

    \begin{claim}
        \label[claim]{claim:qma-h4-to-h5}
        Let $\Pi$ be a perfectly $2t$-private MPQC protocol, then the hybrids $\cH_4$ and $\cH_5$ are perfectly indistinguishable.
    \end{claim}

    \begin{proof}
         This claim directly follows with the observation made in~\cref{lem:t-out-of-n-quantum} that the above described scheme is a $2t$-out-of-$n$ secret sharing scheme and that the security of it remains after applying the Circuit-to-Hamiltonian reduction.\footnote{Otherwise an adversary could simply apply the reduction itself and break the security of the secret sharing scheme.} Therefore, we can rely on the results of Damg\r{a}rd et al.~\cite[Proposition 1]{DamgardFNS13} which state that a perfectly secure $2t$-out-of-$n$ quantum secret sharing scheme is secure against superposition attacks/$F$-share capture attacks as long as only $t/2$ of the shares are revealed. This directly implies the perfect indistinguishability of the hybrids $\cH_4$ and $\cH_5$ since in the interaction between the prover and the verifier in the third round only $t/2$ shares are revealed.
\end{proof}

\end{proof} 

\subsection*{Acknowledgments}

RJ and DK were supported in part by AFOSR, NSF CAREER CNS-2238718, NSF 2112890, NSF CNS-2247727 and a Google Research Scholar award. This material is based upon work supported by the Air Force Office of Scientific Research under award number FA9550-23-1-0543. GM is supported by the European Research Council through an ERC Starting Grant (Grant agreement No.~101077455, ObfusQation) and by the Deutsche Forschungsgemeinschaft (DFG, German Research Foundation) under Germany's Excellence Strategy - EXC 2092 CASA – 390781972.

\printbibliography

\addcontentsline{toc}{section}{References}
\appendix

\section{Proof of Security Against Superposition Attacks~\cite{DamgardFNS13}}
\label[appendix]{app:dfns-thm1}

\begin{theorem}[\cref{thm:dfns}]
Let $S$ be a secret sharing scheme with adversary structure $G$. If $F^2 \subseteq G$ where $F^2 \triangleq \{A \: \vert \: \exists B, C \in F \: :\: A = B \cup C\}$, then $S$ will be perfectly secure against superposition $F$-attacks (\cref{def:ps-sup}).
\end{theorem}

\begin{proof}[Proof of \cref{thm:dfns}]

We consider the adversary's final state for any $s \in \bbS$,
\begin{align*}
    \rho^{adv, F}_s &= \sum_{r \in \cR} p_r \sum_{\substack{x, x' \in \cX \\ a, a' \in \zo^t\\ A, A' \in F}} \alpha_{x',A',a'}^* \: \alpha_{x,A,a} \: \ket{x'}_e\bra{x}_e \otimes \ket{A', a' \oplus v_{A'}(s; r)}_q\bra{A, a \oplus v_{A}(s; r)}_q\\
    &= \sum_{\substack{x, x' \in \cX \\ a, a' \in \zo^t\\ A, A' \in F}} \alpha_{x',A',a'}^* \: \alpha_{x,A,a} \: \ket{x'}_e\bra{x}_e \otimes \sum_{r \in \cR} p_r \ket{A', a' \oplus v_{A'}(s; r)}_q\bra{A, a \oplus v_{A}(s; r)}_q.
\end{align*}
We rewrite our state in the following form
\begin{align}
    \rho^{adv, F}_s &= \sum_{\substack{a, a' \in \zo^t\\ A, A' \in F}}\rho^{adv, F}_s\vert_{a,a',A,A'} \label{eqn:fins}
\end{align}
where we define
\begin{align}
    \rho^{adv, F}_s\vert_{a,a',A,A'} &= \sum_{\substack{x, x' \in \cX}} \alpha_{x',A',a'}^* \: \alpha_{x,A,a} \: \ket{x'}_e\bra{x}_e \otimes \sum_{r \in \cR} p_r \ket{A', a' \oplus v_{A'}(s; r)}_q\bra{A, a \oplus v_{A}(s; r)}_q. \label{eqn:flns1}
\end{align}

Let us fix any arbitrary $a,a' \in \zo^t$, and $A, A' \in F$. 
Temporarily consider the state in Equation~\ref{eqn:flns1} for some secret $s \in \bbS$.
We notice that only the second term depends on the secret $s$, so we focus on this term. This term contains information about the views of parties in $B= A' \cup A \in F^2 \subseteq G$. Since our secret sharing scheme $S$ has adversary structure $G$, $S$ is perfectly secure against classical $G$-attacks which ensures that the distribution of $v_B(s; r)$ does not depend on $s$. 
This, in turn, translates to the statement that for all $s, s' \in \bbS$, 
we will have
\begin{align*}
    \sum_{r \in \cR} p_r \ket{A', a' \oplus v_{A'}(s; r)}_q\bra{A, a \oplus v_{A}(s; r)}_q &= \sum_{r \in \cR} p_r \ket{A', a' \oplus v_{A'}(s'; r)}_q\bra{A, a \oplus v_{A}(s'; r)}_q
\end{align*}
which would then imply that
\begin{align*}
    &\rho^{adv, F}_s\vert_{a,a',A,A'} = \sum_{\substack{x, x' \in \cX}} \alpha_{x',A',a'}^* \: \alpha_{x,A,a} \: \ket{x'}_e\bra{x}_e \otimes \sum_{r \in \cR} p_r \ket{A', a' \oplus v_{A'}(s; r)}_q\bra{A, a \oplus v_{A}(s; r)}_q \\
    &= \sum_{\substack{x, x' \in \cX}} \alpha_{x',A',a'}^* \: \alpha_{x,A,a} \: \ket{x'}_e\bra{x}_e \otimes \sum_{r \in \cR} p_r \ket{A', a' \oplus v_{A'}(s'; r)}_q\bra{A, a \oplus v_{A}(s'; r)}_q = \rho^{adv, F}_{s'}\vert_{a,a',A,A'}.
\end{align*}
Observe that since we fixed arbitrary values for terms $s \in \bbS$, $a, a' \in \zo^t$, and $A,A' \in F$, the above argument will follow for any value. 

Furthermore, by our definition in Equation~\ref{eqn:fins}, we have for all $s, s' \in \bbS$
\begin{align*}
    \rho^{adv, F}_s &= \sum_{\substack{a, a' \in \zo^t\\ A, A' \in F}}\rho^{adv, F}_s\vert_{a,a',A,A'} = \sum_{\substack{a, a' \in \zo^t\\ A, A' \in F}}\rho^{adv, F}_{s'}\vert_{a,a',A,A'} = \rho^{adv, F}_{s'}.
\end{align*}
Thus proving that $S$ will be secure against superposition $F$-attacks.
\end{proof}

\end{document}